\documentclass[fleqn,usenatbib]{mnras}

 \usepackage{newtxtext,newtxmath}

\usepackage[T1]{fontenc}

\DeclareRobustCommand{\VAN}[3]{#2}
\let\VANthebibliography\thebibliography
\def\thebibliography{\DeclareRobustCommand{\VAN}[3]{##3}\VANthebibliography}

\usepackage{graphicx}	
\usepackage{amsmath}	
\usepackage{multirow}
\usepackage{pdflscape}
\usepackage{subcaption}
\usepackage{xcolor}

\title[Filament and H{\sc i} spin orientation]{MIGHTEE-H{\sc i}: The relation between the H{\sc i} gas in galaxies and the cosmic web}

\author[M. N. Tudorache et al.]{Madalina N. Tudorache,$^{1}$\thanks{E-mail: madalina.tudorache@physics.ox.ac.uk}
M. J. Jarvis,$^{1,2}$
I. Heywood,$^{1,3,4}$
A. A. Ponomareva,$^{1}$
N. Maddox,$^{5}$
\newauthor
B. S. Frank,$^{4,6,7}$
N. J. Adams,$^{8}$
R. A. A. Bowler,$^{8}$
I. H. Whittam,$^{1,2}$
M. Baes,$^{9}$
H. Pan,$^{2,1}$
\newauthor
S. H. A. Rajohnson,$^{7}$
F. Sinigaglia,$^{10,11}$
K. Spekkens$^{12}$
\\
$^{1}$Astrophysics, Department of Physics, University of Oxford, Keble Road, Oxford OX1 3RH, UK\\
$^{2}$Department of Physics and Astronomy, University of the Western Cape, Robert Sobukwe Road, 7535 Bellville, Cape Town, South Africa\\
$^{3}$Department of Physics and Electronics, Rhodes University, PO Box 94, Makhanda, 6140, South Africa \\
$^{4}$South African Radio Astronomy Observatory, 2 Fir Street, Black River Park, Observatory, Cape Town 7925, South Africa\\
$^{5}$University Observatory, Faculty of Physics, Ludwig-Maximilians-Universit\"at,
Scheinerstr. 1, 81679 Munich, Germany\\
$^{6}$The Inter-University Institute for Data Intensive Astronomy (IDIA), Department of Astronomy, University of Cape Town, Private Bag X3, Rondebosch, \\ 7701, South Africa \\
$^{7}$Department of Astronomy, University of Cape Town, Private Bag X3, Rondebosch 7701, South Africa \\
$^{8}$Jodrell Bank Centre for Astrophysics, Department of Physics and Astronomy, School of Natural Sciences, The University of Manchester, Manchester, \\ M13 9PL, UK \\
$^{9}$Sterrenkundig Observatorium, Department of Physics and Astronomy, Universiteit Gent, Krijgslaan 281 S9, B-9000 Gent, Belgium \\
$^{10}$Department of Physics and Astronomy, Università degli Studi di Padova, Vicolo dell’Osservatorio 3, I-35122, Padova, Italy \\
$^{11}$INAF - Osservatorio Astronomico di Padova, Vicolo dell’Osservatorio 5, I-35122, Padova, Italy \\
$^{12}$Department of Physics and Space Science, Royal Military College of Canada, PO Box 17000, Station Forces, Kingston, Ontario,
K7K 7B4, Canada
}

\date{Accepted XXX. Received YYY; in original form ZZZ}

\pubyear{2021}

\begin{document}
\label{firstpage}
\pagerange{\pageref{firstpage}--\pageref{lastpage}}
\maketitle

\begin{abstract}

We study the 3D axis of rotation (3D spin) of 77 H{\sc i} galaxies from the MIGHTEE-H{\sc i} Early Science observations, and its relation to the filaments of the cosmic web. For this H{\sc i}-selected sample, the alignment between the spin axis and the closest filament ($\lvert \cos \psi \rvert$) is higher for galaxies closer to the filaments, with $\langle\lvert \cos \psi \rvert\rangle= 0.66 \pm 0.04$ for galaxies $<5$\,Mpc from their closest filament compared to $\langle\lvert \cos \psi \rvert\rangle= 0.37 \pm 0.08$ for galaxies  at $5 < d <10$\,Mpc. We find that galaxies with a low H{\sc i}-to-stellar mass ratio ($\log_{10}(M_{\rm HI}/M_{\star}) < 0.11$) are more aligned with their closest filaments, with $\langle\lvert \cos \psi \rvert\rangle= 0.58 \pm 0.04$; whilst galaxies with ($\log_{10}(M_{\rm HI}/M_{\star}) > 0.11$) tend to be mis-aligned, with $\langle\lvert \cos \psi \rvert\rangle= 0.44 \pm 0.04$. We find tentative evidence that the spin axis of H{\sc i}-selected galaxies tend to be aligned with associated filaments ($d<10$\,Mpc), but this depends on the gas fractions. Galaxies that have accumulated more stellar mass compared to their gas mass tend towards stronger alignment. 
Our results suggest that those galaxies that have accrued high gas fraction with respect to their stellar mass may have had their spin axis alignment with the filament disrupted by a recent gas-rich merger, whereas the spin vector for those galaxies in which the neutral gas has not been strongly replenished through a recent merger tend to orientate towards alignment with the filament. We also investigate the spin transition between galaxies with a high H{\sc i} content and a low H{\sc i} content at a threshold of $M_{\mathrm{H{\sc{I}}}}\approx 10^{9.5} M_{\odot}$ found in simulations, however we find no evidence for such a transition with the current data.
\end{abstract}




\section{Introduction}

On the largest scales, the Universe contains a network-like distribution of galaxies, gas and dark matter - the cosmic web \citep{bond-1996}. This cosmic web, formed of clusters, walls, voids and filaments, was predicted by Zel'dovich's model of the evolution of the non-linear growth of primordial density perturbations \citep{zeldovich-1970}. Early evidence of its existence was presented in \citet{Davis1982} and \citet{deLapparent-1986}, who found a web-like distribution of galaxies. The efforts to trace this elusive web have continued since, as more and more galaxy surveys such as the Sloan Digital Sky Survey (SDSS, \citealt{sdss}), the 2dF Galaxy Redshift Survey (2dFGRS, \citealt{2dfgrs}) or the 2MASS Redshift Survey (2MRS, \citealt{2mass}) were conducted. As well as using galaxy catalogues, N-body simulations showed that cold dark matter (CDM) could be responsible for the formation of the voids, walls and filaments \citep[e.g.][]{springel_2005}.

Tracing the cosmic web is vital for understanding the environment of the galaxies. Several galaxy properties, such as stellar mass, colour, star formation rate (SFR) and specific star formation rate (sSFR) have been shown to be sensitive to the external conditions. Using the SDSS, \citet{Kuutma_2017} showed that for a fixed environment density level, there was a higher elliptical-to-spiral ratio towards the filament spines and a decrease in SFR, however this occurs without an increase in mass. \citet{Laigle_2017} showed that for the COSMOS catalogue, massive galaxies were closer to filaments, and for fixed stellar mass, red galaxies were closer to the spine of the filament. Furthermore, studies such as \citet{kleiner} and \citet{crone-odekon} have investigated the link between the neutral hydrogen (H{\sc i}) content in galaxies and the large-scale structures, with different results regarding the correlation between position of the galaxy and its H{\sc i} content, and how it is fuelled by the filaments.

A key property of galaxies is their angular momentum, which could improve our understanding of their morphology and its dependence on the environment. Strong evidence of the alignment of the angular momentum vector of the galaxies and their associated filament is yet to be found. Since the first study on this topic in a hydrodynamical simulation by \citet{hahn}, who reported that massive disk galaxies were aligned with the filaments, several other hydrodynamical simulation studies carried out do not agree with this result. The most significant prediction in the current literature is that low-mass galaxies tend to be rotationally aligned with their closest filaments, whilst high-mass galaxies have a tendency towards mis-alignment. This arises from the theoretical considerations of the tidal torque theory \citep{peebles, white}, which relates the spin angular momentum of a proto-galaxy to its tidal interactions with the surrounding matter. Simulations such as those by \citet{aragon-calvo_2007}, \citet{dubois,codis} and \citet{Kraljic_2020} confirm this result, whilst \citet{ganeshaiah-veena-2018} also predict alignment between the angular momentum vector of the dark matter haloes and the filaments of the cosmic web. Whilst the simulations mentioned above find hints of a spin transition from alignment to mis-alignment - for example, \citet{Kraljic_2020} find a spin transition for a stellar mass of $\sim 10^{10} M_{\odot}$, other simulations such as \citet{ganeshaiah-veena-2018} find no such transition and report a preference for overall mis-alignment at all masses. 

Observational evidence for a spin-alignment at a certain stellar mass is lacking. \citet{Krolewski_2019} reports no spin alignment using the Mapping Nearby Galaxies at Apache Point Observatory \citep[MaNGA;][]{manga} integral-field survey, whilst \citet{welker} finds a spin transition within a stellar-mass interval of $10^{10.4} M_{\odot} - 10^{10.9} M_{\odot}$ using the Sydney-AAO (Australian Astronomical Observatory) Multi-object survey (SAMI, \citealt{sami}). 

Concentrating on the H{\sc i} gas, \citet{Kraljic_2020} find a possible spin transition threshold in H{\sc i} mass at $M_{\mathrm{H{\sc{I}}}} = 10^{9.5} M_{\odot}$ using the SIMBA \citep{simba-sim} simulation. \citet{bird2019chiles}, using the COSMOS H{\sc i} Large Extragalactic Survey (CHILES, \citealt{chiles}), find that the spins of their galaxies in their HI-selected sample tend to be aligned with the cosmic web. However, their study does not find any significant mass transition between the aligned and the mis-aligned spin. In addition to a mass dependence, the type of the galaxy has also been shown to relate to the filaments. \citet{Kraljic_2021} using the MaNGA integral-field survey find that the spins of late-type galaxies (LTGs) are preferentially aligned to their closest filament, whilst S0 type galaxies have a preferential perpendicular alignment to their closest filament. The result regarding the elliptical/S0 galaxies has been previously identified in studies such as \citet{tempel2013} using SDSS and \citet{pahwa} using the 2MASS Redshift Survey. Scd types have also been shown to have a preferential parallel alignment, whilst Sab galaxies have been shown to have a preferential perpendicular alignment \citep{hirv}. There are, however, studies which find no particular preference for alignment for spiral galaxies \citep{pahwa, Krolewski_2019}, or they find a perpendicular preference for alignment for these galaxies \citep{Lee_2007}.

As is clear, the overall picture of the relationship between the galaxy spin vector and the direction of the filaments in which it may reside is confusing, with many results from both simulations and observations disagreeing to different degrees. Thus, further work is necessary to elucidate the link between these properties of galaxies and the large scale structure.

For this study, we use a H{\sc i} galaxy sample provided by the MeerKAT International GigaHertz Tiered Extragalactic Exploration (MIGHTEE,  \citealt{mightee}) Early Science release to compute the 3D spin vector of the galaxies, using optically-selected galaxies from the COSMOS and XMM-LSS fields, to find the filaments of the cosmic web. The sample we use is the largest H{\sc i} sample to date used to conduct a study like this, which enables us to make stronger statistical statements than in \citet{bird2019chiles}. This allows us to compute the angular momentum using position angles and inclinations from the H{\sc i} moments, which was not possible in studies such as \citet{kleiner} or \citet{crone-odekon}, since they used single-dish surveys with large numbers of detections, but without the power to sufficiently resolve kinematics for many of the sources. The structure of this paper is organised as follows. Section \ref{sec:method} introduces the MIGHTEE Survey, as well as the methods employed to compute the cosmic web and the spin of the H{\sc i} galaxies. Section \ref{sec:results} discusses the results obtained. The summary and conclusions are presented in Section \ref{sec:conclusions}.

\section{Methodology}
\label{sec:method}

\begin{table}
\centering
\caption{Short summary of the MIGHTEE-HI Early Science
  data products used in this paper.}
\label{tab:mightee}
\begin{tabular}{p{3.5cm} p{3.5cm}}
\hline\hline
\multirow{2}{*}{Area covered} & $\sim 1$\,deg$^2$ COSMOS field \\
                  &  $ \sim 3$\,deg$^2$ XMM-LSS field  \\
Frequency range   &  $ 1320 - 1410 $\,MHz \\
Redshift range   &  $ 0.02 - 0.09 $ \\
Channel width     &  $209$\,kHz \\
Median $\mathrm{H}_{\mathrm{I}}$ channel rms noise  &  $85$ $\mu$Jy \,beam$^{-1}$\\
$N_{\mathrm{HI}}$ sensitivity ($3 \sigma$)   & $1.6 \times 10^{20}$ \,cm$^{-2}$ (per channel) \\ 
\multirow{2}{*}{Synthesised beam} & $14.5" x 11"$ COSMOS field \\
                                 & $12" x 10"$ XMM-LSS field \\
\multirow{2}{*}{H{\sc i} mass lower limit} & $\sim10^{6.7} M_{\odot}(z = 0.02)$ \\
                                    & $\sim 10^{8.5} M_{\odot} (z = 0.09)$\\ 
\hline \hline
\end{tabular}
\end{table}

\subsection{The MIGHTEE survey}
\label{sec:mightee}

\begin{figure*}
  	\includegraphics[width=1\textwidth]{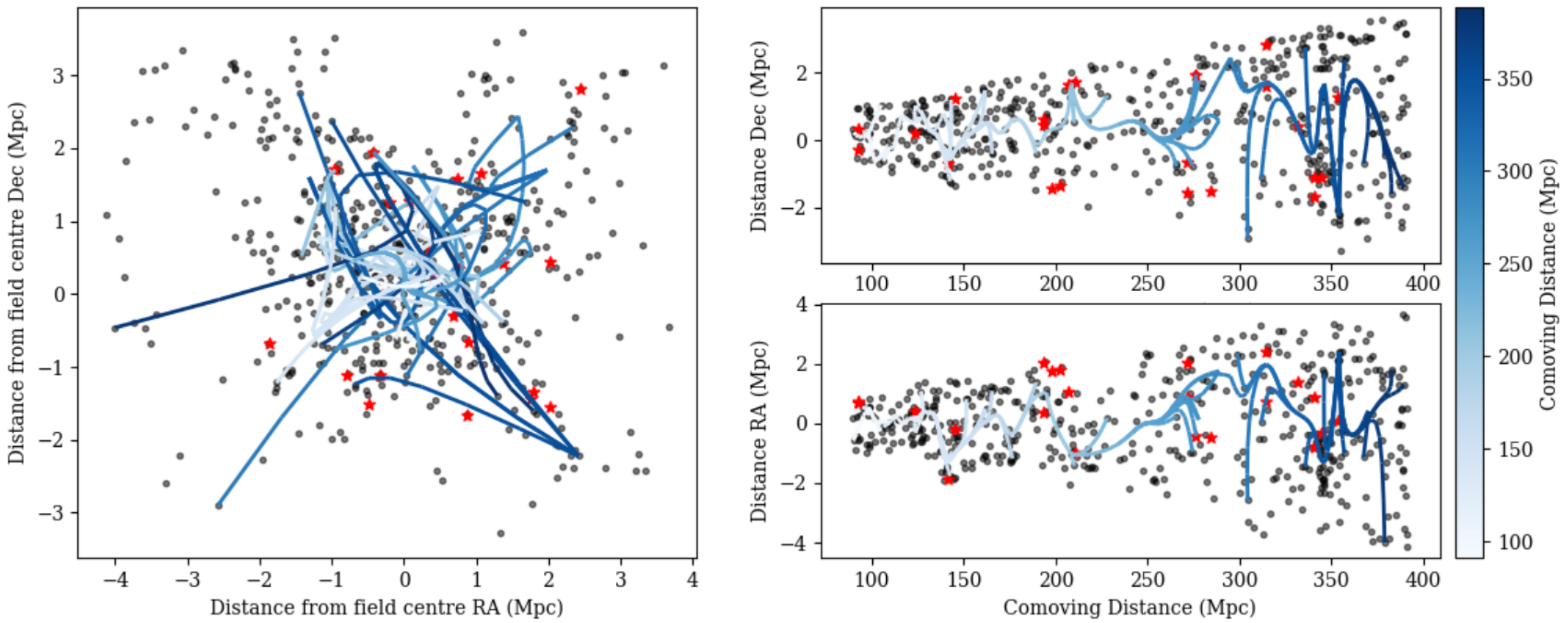}
    \caption{The filament distribution projected in 2D obtained by running DisPerSE with mirror boundary conditions for the COSMOS catalogue in a redshift interval $0.02 < z < 0.09$. Left: Angular distance in right ascension versus angular distance in declination. Top right: Radial comoving distance versus angular distance in declination. Bottom right: Radial comoving distance versus angular distance in right ascension of the filament distribution overlaid on top of the galaxies in the optical sample. The red stars represent the H{\sc i} galaxies detected by MIGHTEE. The colour bar represents the radial comoving distance in Mpc.}
    \label{fig:filament-network}
\end{figure*}

The MIGHTEE survey is one of the eight Large Survey Projects (LSPs) which are being undertaken by MeerKAT \citep{meerkat}. MeerKAT consists of an array of 64 offset-Gregorian dishes, where each dish consists of a main reflector with a diameter of $13.5$\,m and a sub-reflector with a diameter of $3.8$\,m. MeerKAT's three band receivers, UHF–band ($580 < \nu < 1015 $\,MHz), L--band ($900 < \nu < 1670$\,MHz) and S–band ($1750 < \nu < 3500$\,MHz) all collect data in spectral mode. The MIGHTEE survey has three major components: radio continuum \citep{Heywood2021}, polarisation and spectral line.
For this work we use the spectral line information in the L--band with $4096$ channels with a channel width of $209$\,kHz, which corresponds to $44$\,km\,s$^{-1}$ at $z=0$.

MIGHTEE-H{\sc i} \citep{Maddox_2021} is the H{\sc i} emission part of the MIGHTEE survey. Its initial data products, as part of the Early Science release, were obtained using the \texttt{ProcessMeerKAT} calibration pipeline (Frank et. al in prep.). This pipeline is a paralellised \texttt{CASA}\footnote{\url{http://casa.nrao.edu}}-based \citep{casa} pipeline whose calibration routines and strategies are standard (i.e.  flagging, delay, bandpass, and complex gain calibration). It performs spectral-line imaging using \texttt{CASA}’s \texttt{TCLEAN} task. The continuum subtraction was done in two domains. Visibility  domain subtraction was perfomed using the standard \texttt{CASA} routines \texttt{UVSUB} and \texttt{UVCONTSUB}. This process was followed by an image plane based continuum subtraction using per-pixel median filtering, which was applied to the resulting data cubes to reduce the impact of the direction-dependent artefacts. An in-depth description of the procedures employed for data reduction and data quality assessment will be presented in Frank et al. (in prep.). The summary of the data used in this paper is shown in Table \ref{tab:mightee}.

There are $\sim 270$ galaxies in the full Early Science H{\sc i} catalogue. In this paper we use a reduced sample of $77$ galaxies taken from \citet{Ponomareva2021}, with the number being lower than the full Early Science catalogue due to two factors. The first is that we could not obtain accurate kinematically measured inclination and position angles for 183 galaxies due to insufficient signal-to-noise (our kinematic modelling \citep{Ponomareva2021} requires $>3.5\sigma$ per resolution element) and/or not being sufficiently spatially or spectrally resolved in the MIGHTEE data cube (our kinematic modelling requires at least three resolution elements, both spatially and spectrally; beam and channel width are shown in Table \ref{tab:mightee}). The second factor is that the redshift range we chose ($0.02 < z < 0.09$) means we remove a further $10$ galaxies from the sample. We adopt this cut due to the small number of spectroscopic redshifts for galaxies within the COSMOS and XMM-LSS fields at $z < 0.02$, which do not provide enough information to identify the filamentary structures.

\subsection{The Cosmic Web}
\label{sec:cosmic-web}

Several computational techniques have been developed for the detection of large-scale structures (see \cite{Libeskind_2017} for a full review of all available algorithms). The Discrete Persistent Structure Extractor (DisPerSE, \citealt{sousbie}) is a topological algorithm based on discrete Morse theory \citep{milnor1963morse} that computes the skeleton of the cosmic web using Delaunay Tessellation Field Estimator \citep[DTFE;][]{dtfe}. This is achieved by using the DTFE on a distribution of particles or a distribution of galaxies resulting in a density field. As the filaments are string-like structures connecting the galaxy clusters and bordering the voids, the DTFE can easily detect the variations in the field due to the structures. DisPerSE has been used to compute the filaments in studies such as \citet{Galarraga_Espinosa_2020} for the Illustris-TNG simulation \citep{Nelson_2019}, in \citet{Laigle_2017} for the Horizon-AGN simulation \citep{horizon-agn} and in \citet{Kraljic_2020} for the SIMBA simulation \citep{simba}. DisPerSe has also been used to compute filaments from observations such as the CHILES survey by \citet{luber}, the Galaxy And Mass Assembly survey (GAMA, \citealt{gama}) by \citet{Kraljic_2018} and SDSS by \citet{Winkel_2021}.
 
In this paper we use DisPerSE to determine the skeleton of the cosmic web based on the distribution of galaxies from the COSMOS and XMM-LSS fields. These are two of the most widely studied extragalactic fields accessible from the southern hemisphere and have been the subject of a large number of multi-wavelength surveys over the past decade. In this paper we use the imaging data described in \cite{adams2021evolution}, which includes optical and near-infrared imaging from the HyperSuprimeCam Strategic Survey Programme
DR1 \citep[HSC; ][]{Aihara2018} and near-infrared imaging is sourced from the UltraVISTA survey in the COSMOS field \citep{McCracken2012} and the VISTA Deep Extragalactic Observations \citep[VIDEO; ][]{Jarvis2013} Survey in the XMM-LSS field. Spectroscopic redshifts from a variety of surveys have been compiled by the HSC team\footnote{\url{ https://hsc-release.mtk.nao.ac.jp/doc/index.
php/dr1_specz/}} and we use these in this paper.
These spectroscopic data provide the accurate redshifts for the filament finding and the imaging data provide the spectral baseline to derive stellar mass estimates of the galaxies within our sample.

We use all spectroscopic redshifts over the redshift range $0.02<z<0.09$, from a range of surveys and summarised in \cite{adams2021evolution}. Within this range, we have 500 spectroscopic redshifts in the COSMOS field, and 2197 spectroscopic redshifts in the XMM-LSS field. We note that the heterogeneous nature of the spectroscopic redshifts across these fields may results in biases in some relations. For this reason, we restrict our analysis to those relations that should be largely invariant to the heterogeneous nature of the spectroscopic redshifts, i.e. those concerning the alignment of the galaxies' spin axis with the direction of the filaments. 

To find the filaments, the critical points of the density field are identified: we obtain maxima, minima and saddle points. The filaments themselves are computed by connecting a maximum point to a saddle point. As an algorithm, DisPerSE can also return the walls, voids and clusters if needed. The distribution of the filaments is dependent on two important parameters: the boundary conditions (BCs) and the significance level. DisPerSE can be used with four boundary conditions, which deal with the edges of the distribution: periodic, mirror, void and smooth. The periodic boundary conditions consist of normal periodic conditions. The mirror boundary conditions are implemented by generating galaxies which mirror the ones on the edges, whilst the smooth boundary conditions are implemented by generating galaxies based on an interpolated density field. The void boundary conditions, which do not add any boundary galaxies at the edges, fail to provide a coherent filament network - so it was discarded in our work. 

The other parameter, the significance level, deals with the level at which a structure is picked up as a filament or not. The significance level is described by a threshold in a persistance diagram - which can be drawn as the absolute value of the ratio between two critical points in a pair and the value at the lowest critical point of the two (the background density). The choice of threshold heavily influences the detected filaments, since a low threshold will pick out structures that might not be filaments, whilst a high threshold might smooth out real structures.

Following \citet{luber}, we split our sample into slices of equal redshifts $\Delta z = 0.01$ from $z=0.02$ to $z=0.09$. We choose a threshold of $3.5 \sigma$, motivated by the comparisons in \citet{sousbie} and use the mirror boundary conditions to aid comparison with previous work \citep[e.g.][]{bird2019chiles}. Figure \ref{fig:filament-network} shows an example for the filament network converted into 3D Cartesian coordinates for mirror boundary conditions along with the galaxies used to compute it. The choice of the significance level ensures that our filaments are robust and that the big structures are picked up without washing out some of the relevant finer structures.

\begin{figure}
	\includegraphics[width=\columnwidth]{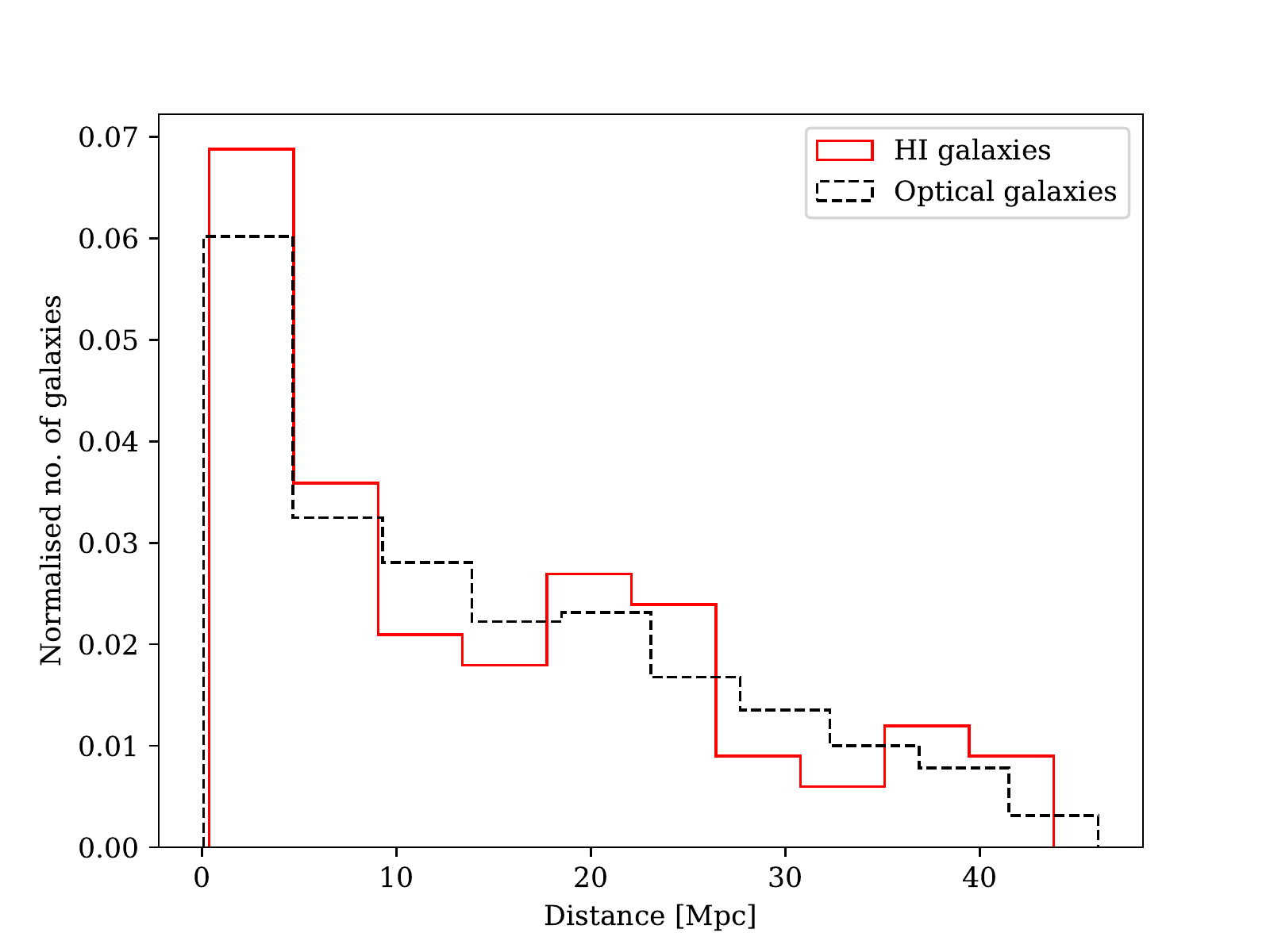}
    \caption{Normalised histogram of both the H{\sc i} galaxies (red solid line) and the optical galaxies (black dashed line) from the COSMOS and XMM-LSS fields in the redshift range of $ 0.02 < $ z $ < 0.09$ as a function of distance from the cosmic web obtained using mirror boundary conditions on DisPerSE.}
    \label{fig:hi-dist}
\end{figure}

\begin{figure*}
\begin{subfigure}[b]{.49\textwidth}
  \centering
  \captionsetup{justification=centering}
  \includegraphics[width=0.95\linewidth]{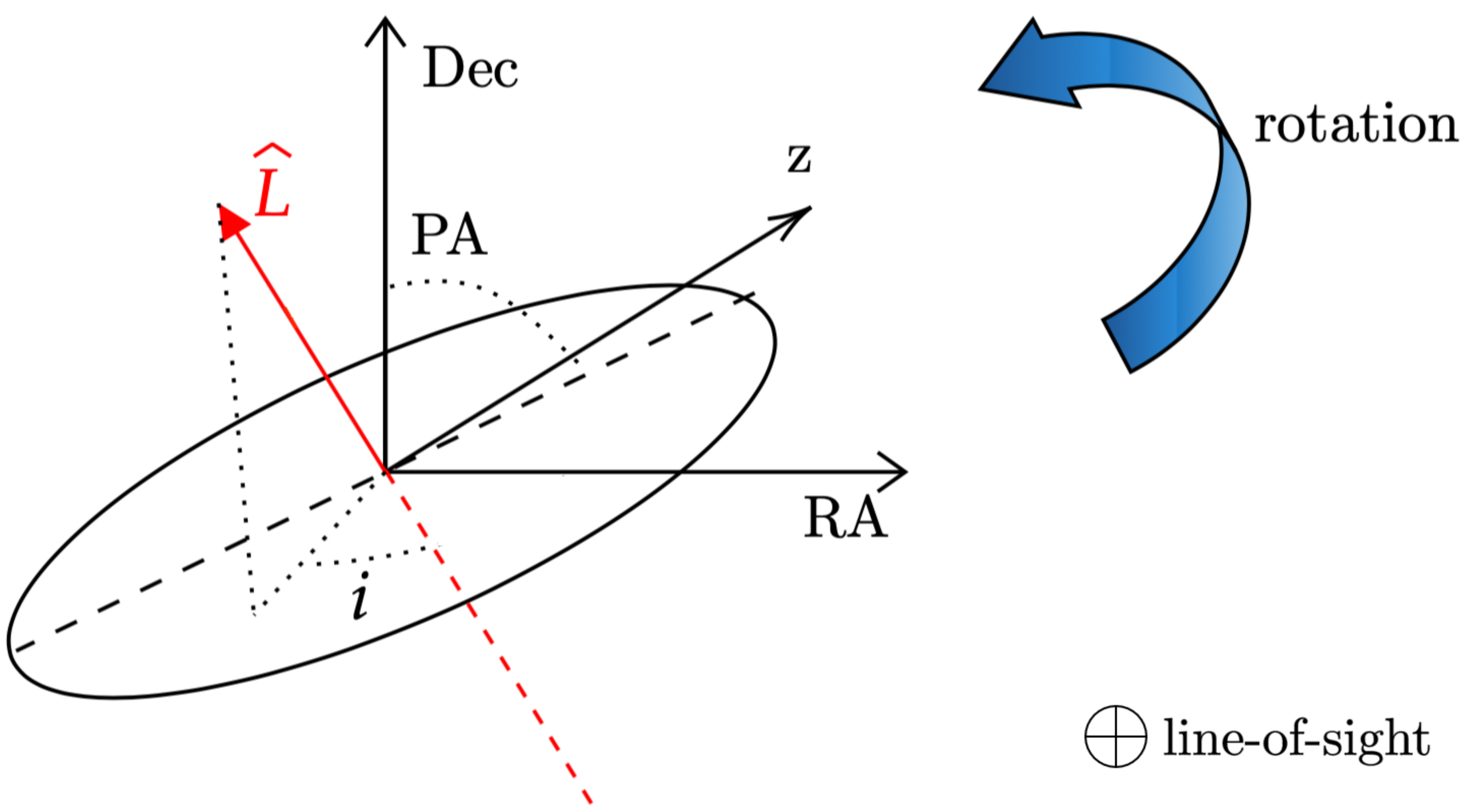} 
  \caption{}
  \label{subfig:galaxy-diagram}
\end{subfigure}
\begin{subfigure}[b]{.49\textwidth}
  \centering
  \captionsetup{justification=centering}
  \includegraphics[width=1\linewidth]{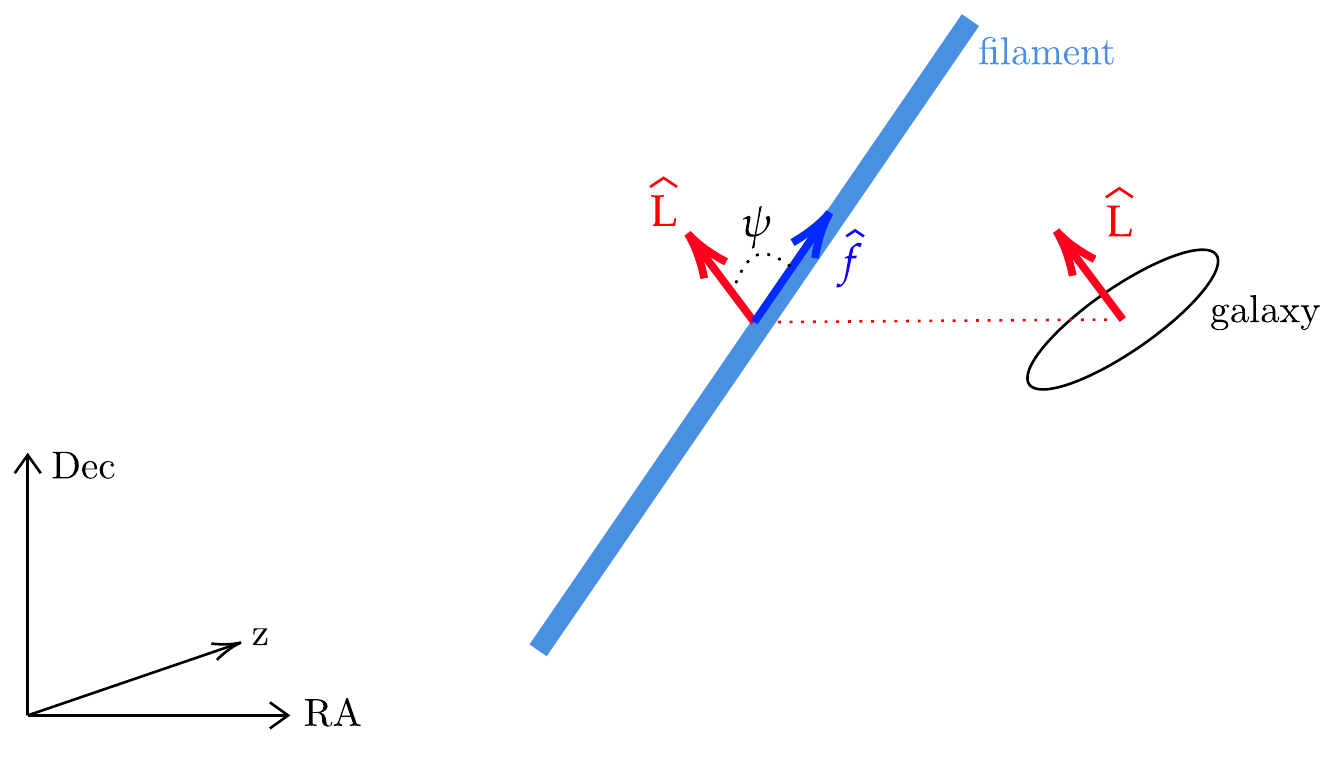}
  \caption{}
  \label{subfig:psi-diagram}
\end{subfigure}
\caption{Schematic of a galaxy and its closest filament in (RA, Dec, z) coordinates. ({\it left}) Illustration of the position angle (PA), the inclination ($i$), the unit spin vector ($\hat{L}$) and the direction of rotation for a galaxy. ({\it right}) Illustration of a galaxy and its unit spin vector next to its closest filament vector along with the angle $\psi$ between them.}
  \label{fig:angles-diagram}
\end{figure*}

\subsection{Distance to the closest filament}
\label{sec:distance-filament}

To calculate the distance from a galaxy to the spine of the closest filament, we used the skeleton generated by DisPerSE - which consists of a network of points which can be used to construct small segments to assemble the filament network. The midpoint of each segment was then cross-matched with our H{\sc i} galaxy sample in each redshift slice. This provides the physical separation - for which we use the 3D position of the filament point - between each galaxy and its closest filament. This method does not provide the true exact distance to the filament, as it would require computing the right angle between the point and the segment. However, the size of  filament segments are on average $\sim 1$\,Mpc. In the extreme case of the galaxy lying within 1\,Mpc of the filament this would result in an uncertainty of the order unity, however the fractional uncertainty obviously decreases for a galaxy that resides at a greater distance to the filament: for a  galaxy 5\,Mpc from the filament, the uncertainty on the distance is $\sim 1$ per cent due to this assumption, i.e. a negligible source of uncertainty given the uncertainty around the filament distribution itself.
Figure \ref{fig:hi-dist} shows the histogram of the distances from the H{\sc i} galaxies in our sample and the nearest filament (red), as well as the galaxies with spectroscopic redshifts from the optical catalogue used to compute the large-scale structures and their closest filament (black). As can be seen, the galaxies are all $ \lesssim 50$\,Mpc from the nearest filaments (and $\lesssim 40$\,Mpc for the H{\sc i} galaxies). When measuring the 3D distance, we did not take into account redshift distortion effects as we considered it a $2^{\mathrm{nd}}$ order effect.

\subsection{Spin of the galaxy}
\label{sec:hi-spin}

Following the treatment in \citet{Lee_2007}, which uses a thin-disk approximation, the spin unit vector of a galaxy can be characterised in local spherical coordinates as:
\begin{align}
    \hat{L}_{r} &=\cos i \\
    \hat{L}_{\theta} &= \sin i \sin \mathrm{PA} \\
    \hat{L}_{\phi} &= \sin i \cos \mathrm{PA}
    \label{eq:hi-spin-sph}
\end{align}
where PA is the position angle and $i$ is the inclination angle of the galaxy. The values for both the PAs and the $i$'s were measured using 3D kinematic modelling, as presented in \citet{Ponomareva2021}.
The inclination angle is defined as $i = 0$ if face-on and $i = \pi/2$ if edge-on, whilst the PA is measured from the north counterclockwise to the receding side of a galaxy.

The unit spin vector is converted into Cartesian coordinates, with the spherical vector related to the Cartesian vector by:
\begin{equation}
\left[\begin{array}{c}
\hat{L}_{x} \\
\hat{L}_{y} \\
\hat{L}_{z}
\end{array}\right]=\left[\begin{array}{ccc}
\sin \alpha \cos \beta & \cos \alpha \cos \beta & -\sin \beta \\
\sin \alpha \sin \beta & \cos \alpha \sin \beta & \cos \beta \\
\cos \alpha & -\sin \alpha & 0
\end{array}\right]\left[\begin{array}{c}
\hat{L}_{r} \\
\hat{L}_{\theta} \\
\hat{L}_{\phi}
\end{array}\right] ,
\label{eq:hi-spin-xyz}
\end{equation}
where $\alpha=\pi / 2-\mathrm{DEC}$ and $\beta=\mathrm{RA}$, with DEC and RA corresponding to declination and right ascension, respectively.
Figure \ref{subfig:galaxy-diagram} illustrates the unit spin vector along with the angles used for its calculation and the direction of rotation for the galaxy. There is a sign ambiguity which arises in $\hat{L}_{r}$, which has been shown in \citet{Trujillo_2006}. Following past work \citep{Lee_2007, Kraljic_2021}, we choose to take the positive sign in $\hat{L}_{r}$. \citet{Kraljic_2021} has shown that if the sign of $\hat{L}_{r}$ is flipped in Equation \ref{eq:hi-spin-xyz}, the overall effect of a galaxy being either aligned or mis-aligned does not change due to the symmetry.

\subsection{Angle between galaxy and filament}
\label{sec:cos-angle}

To calculate the angle between the spin galaxy vector and the filament, we crossmatch the galaxy with the closest filament, which is defined by a starting point $f_1(\mathrm{RA}_1, \mathrm{Dec}_1, z_1)$ and an end point $f_2(\mathrm{RA}_2, \mathrm{Dec}_2, z_2)$, where $z$ is the redshift centred on the midpoint of the filament segment. These points are all generated using the skeleton from DisPerSE. To calculate the spherical components of the filament vector we then write:
\begin{align}
    f_{\mathrm{RA}} &= f_2(\mathrm{RA}_2) - f_1(\mathrm{RA}_1) ,\\
    f_{\mathrm{Dec}} &= f_2(\mathrm{Dec}_2) - f_1(\mathrm{Dec}_1) ,\\
    f_{z} &= f_2(z_2) - f_1(z_1).
    \label{eq:filament}
\end{align}

The filament vector is then converted into Cartesian coordinates in order to compute the dot product between the spin vector of the galaxy $\mathbf{L}$ and the filament vector $\mathbf{f}$. To find the cosine of the angle between the galaxy spin vector and the filament vector $\psi$, we divide the dot product by the modulus of the filament vector, as the spin vector is already normalised to 1:
\begin{equation}
    \cos \psi = \frac{f_x \cdot \hat{L}_x + f_y \cdot \hat{L}_y + f_z \cdot \hat{L}_z}{|\mathbf{f}|},
\end{equation}
where $f_x$, $f_y$, $f_z$ are the Cartesian components of the filament vector $\mathbf{f}$ and $\hat{L}_x$, $\hat{L}_y$, $\hat{L}_z$ are defined as before. Figure \ref{subfig:psi-diagram} shows a schematic of the two vectors and $\psi$. To analyse the orientation of the galaxy spin relative to the spine of the filament we take the absolute value of the cosine, which gives us the acute value of the angle $\psi$ independent of the direction of the normalised filament vector $\hat{f}$. Following the convention in \citet{Kraljic_2020}, for $\lvert\cos \psi \rvert < 0.5$ the two are considered mis-aligned, whilst for $\lvert\cos \psi \rvert > 0.5$ they are considered aligned.

\section{Results}
\label{sec:results}

\subsection{Spin alignment as a function of distance-to-filament}
\label{sec:spin-fil-dist}

\begin{figure}
	\includegraphics[width=\columnwidth]{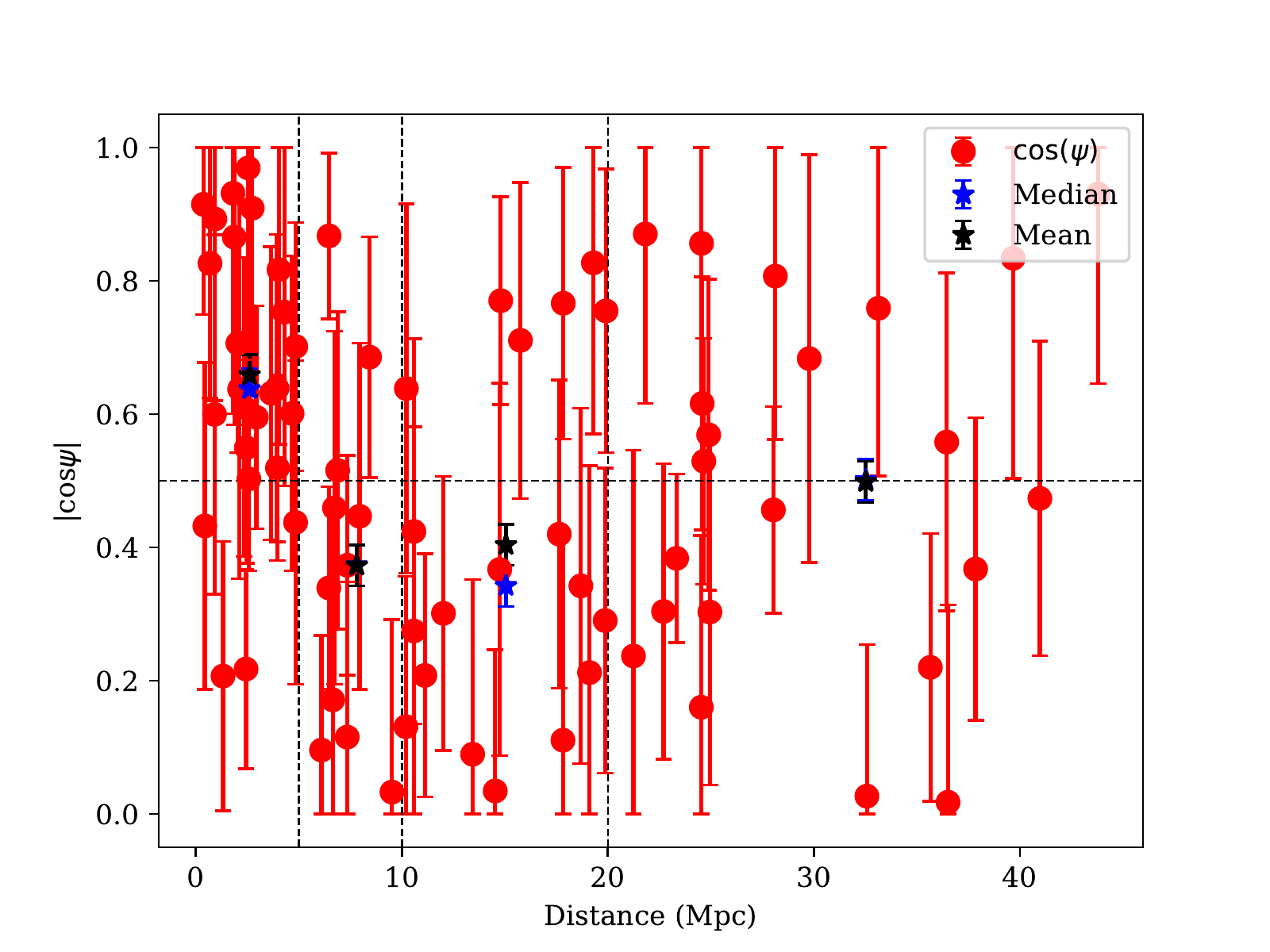}
    \caption{$\lvert \cos \psi \rvert$ for the galaxy sample as a function of distance to the closest filament computed with mirror boundary conditions. The black horizontal dotted line represents the spin value 0.5, whilst the vertical dotted lines mark the 5\,Mpc, 10\,Mpc and 20\,Mpc distance cuts respectively. The blue and black stars represent the medians and the means of each bin, respectively. Uncertainties on the individual values of $\cos \psi$ are determined using the method outlined in Section~\ref{sec:spin-fil-dist}.}
    \label{fig:hi-p-dist}
\end{figure}

As can be seen in Figure~\ref{fig:hi-dist}, all of the H{\sc i}-selected galaxies in our sample lie within $40$\,Mpc of their associated closest filament - with the majority of the sample within 10\,Mpc  ($36$ H{\sc i}-selected galaxies). For our H{\sc i}-selected galaxy sample, $50$\% of the galaxies are within $10.6$\,Mpc, whilst $50$\% of the optical galaxies used to compute the cosmic web are within $11.6$\,Mpc. Using the method described in Section~\ref{sec:cos-angle}, we calculated the spin axis for the 77 galaxies in our sample and the cosine of the angle between the spin axis of each galaxy and its closest filament. The results can be seen in Table \ref{table:cos-dist-m} and Figure \ref{fig:hi-p-dist}, where we separate our sample into different distance ranges. We use these distance ranges as bins for which we calculate the means and the medians, to highlight the overall trend. Whilst the errors on the means (medians) are computed using the standard errors, it is more difficult to  calculate the errors on the individual $\langle\lvert \cos \psi \rvert\rangle$ values. This is due to the nature of the filament-finding algorithm which does not provide any information on the uncertainty in the length or direction of the filaments. Hence, in order to be able to provide an estimate of the uncertainty for $\cos \psi$, we determined the filament distribution  by randomly omitting 5 per cent of the optical galaxies and computing a new network 100 times. We then used the new network to find the closest filament to the galaxy and to recalculate $\cos(\psi)$. Therefore, for each galaxy, we had 101 values for the cos-angle, which allows us to determine the standard deviation on each $\cos(\psi)$ value. These are shown in Figure~\ref{fig:hi-p-dist}. We truncate the error bars where the formal uncertainty gives a value above $\lvert \cos \psi\rvert = 1$ and below $\lvert \cos \psi \rvert = 0$. Clearly the result is noisier where we lack large numbers at the larger distances to the filaments, however there is a clear result for alignment for the galaxies within $5$\,Mpc of their closest filament. 

We use the Mann-Whitney U test \citep{mwu} on the distributions with different distance ranges to compare their medians, shown in Table~\ref{table:cos-dist-m-mwu}. We find significant evidence for alignment between the spin axis of the H{\sc i}-selected galaxies and their closest identified filaments. 
For example, for the galaxies within $5$\,Mpc of their closest filament we find $\langle\lvert \cos \psi \rvert\rangle= 0.66 \pm 0.04$, whereas those galaxies that lie $>5$\,Mpc away from their nearest filament give a mean alignment of $\langle\lvert \cos \psi \rvert\rangle= 0.44 \pm 0.04$, with a Mann-Whitney test giving a $p-$value of $5\times 10^{-4}$, providing strong evidence that the two distributions are significantly different. The trend continues for the distributions of the galaxies within $10$\,Mpc of their closest filament and away from $10$\,Mpc of their nearest filament, with a Mann-Whitney $p-$value of $3.1 \cdot 10^{-2}$. The only distribution without a statistically significant $p-$value is for the galaxies within $20$\,Mpc of their closest filaments and those beyond $20$\,Mpc. This is likely due to the low number of galaxies which are beyond $20$\,Mpc from their closest filament.

This is in agreement with the results obtained by \citet{bird2019chiles}, who used only 10 galaxies (which were within 10\,Mpc of their closest filament) and also found that galaxy spins tend to be aligned with the filaments of the cosmic web.

\begin{table}
	\centering
	\caption{The mean of the cosine of the angle between the H{\sc i} spin of the galaxy and its closest filament $\cos{\psi}$ and the p-values for the Kolmogorov-Smirnov test for different distance ranges. }
	\label{table:cos-dist-m}
	\begin{tabular}{c|c|c|c}
\hline\hline
\textbf{Distance Cut} & ${\langle \lvert\mathbf{cos\psi}\rvert\rangle} $ & $\mathbf{p_{KS}}$ \\ \hline\hline 
$ 0$\,Mpc $ < d < 5 $\,Mpc & $0.66 \pm 0.04$ & $5 \cdot 10^{-2}$ \\ 
$ 5$\,Mpc $ < d < 10$\,Mpc & $0.37 \pm 0.08$ & $ 9 \cdot 10^{-2}$ \\ 
$ 10$\,Mpc $ < d < 20$\,Mpc & $0.40 \pm 0.06$ & $ 4 \cdot 10^{-3}$\\ 
$ d > 20$\,Mpc & $0.50 \pm 0.06$ & $10^{-9}$ \\ 
full range  &  $0.51 \pm 0.03$    &  $10^{-19}$ \\ \hline
\end{tabular}
\end{table}

To verify our result, we shuffled the PAs and the $i$'s in the sample. With this shuffled galaxy sample, we cross-matched it with the filamentary structures and then recalculated the cosine of the angles. We repeated this process 2000 times and used a Kolmogorov-Smirnov test \citep{ks-two} to determine whether our measured alignments are consistent with the null-hypothesis of a randomly oriented spin vector for our galaxy sample, which we show in Table \ref{table:cos-dist-m}.  The $p-$values for the whole sample for the KS test is $10^{-19}$. Thus we can strongly reject the null hypothesis of the spin axis of galaxies being randomly oriented with respect to the orientation of the filaments. 
\begin{table}
	\centering
	\caption{A comparison between the $\cos{\psi}$ distributions for different distance cuts using the p-values for the Mann-Whitney U test.}
	\label{table:cos-dist-m-mwu}
    \begin{tabular}{ccc}
    \hline \hline
    \textbf{Distribution 1}   & \textbf{Distribution 2}     & $\mathbf{p_{MW}}$                                    \\ \hline \hline
    $ d < 5 $\,Mpc     & $ d > 5 $\,Mpc             & $ 10^{-4}$ \\
    $ d < 5 $\,Mpc        & $5$\,Mpc $ < d < 10$\,Mpc   & $ 2 \cdot 10^{-3}$                                 \\
    $d < 10$\,Mpc & $ d > 10$\,Mpc              & $ 3.1 \cdot 10^{-2}$                                  \\
    $ d < 10$\,Mpc & $ 10$\,Mpc $ < d < 20$\,Mpc & $ 1.7 \cdot 10^{-2}$                                 \\
    $ d < 20$\,Mpc            & $ d > 20$\,Mpc              & $ 4.3 \cdot 10^{-1}$                                 \\ \hline
    \end{tabular}
\end{table}

There are not many studies investigating the link between the distance of the galaxy from the filament and $\lvert \cos \psi \rvert$. \citet{Krolewski_2019} do not find any connection between the distance to the filament and the spin of galaxies using the MaNGA integral-field survey. However, the distance cuts they use are much smaller than the ones used in this paper: $0.3$\,Mpc, $1.0$\,Mpc and $1.8$\,Mpc. We can explore larger ranges in our study due to MeerKAT's  field of view. The mean of $\lvert \cos \psi \rvert$ for all their cuts is in the range $\langle\lvert \cos \psi \rvert\rangle= 0.62$ to $\langle\lvert \cos \psi \rvert\rangle= 0.67$, which is in accordance with our results, since for the closest galaxies to the filaments, the alignment is stronger. A key difference in methodology is that they use 2D angles, which could cause a confusion between filaments and walls, leading to the walls dominating the effect \citep{welker}.

\subsection{Spin alignment as a function of H{\sc i} Mass}
\label{sec:hi-mass}

A better understanding of the H{\sc i} content of a galaxy is vital in understanding galaxy evolution. As H{\sc i} extends to larger radii than stars in galaxies, it is more easily perturbed during tidal interactions and hence, more sensitive to external influences \citep{1994Natur.372..530Y}.

\begin{figure}
	\includegraphics[width=\columnwidth]{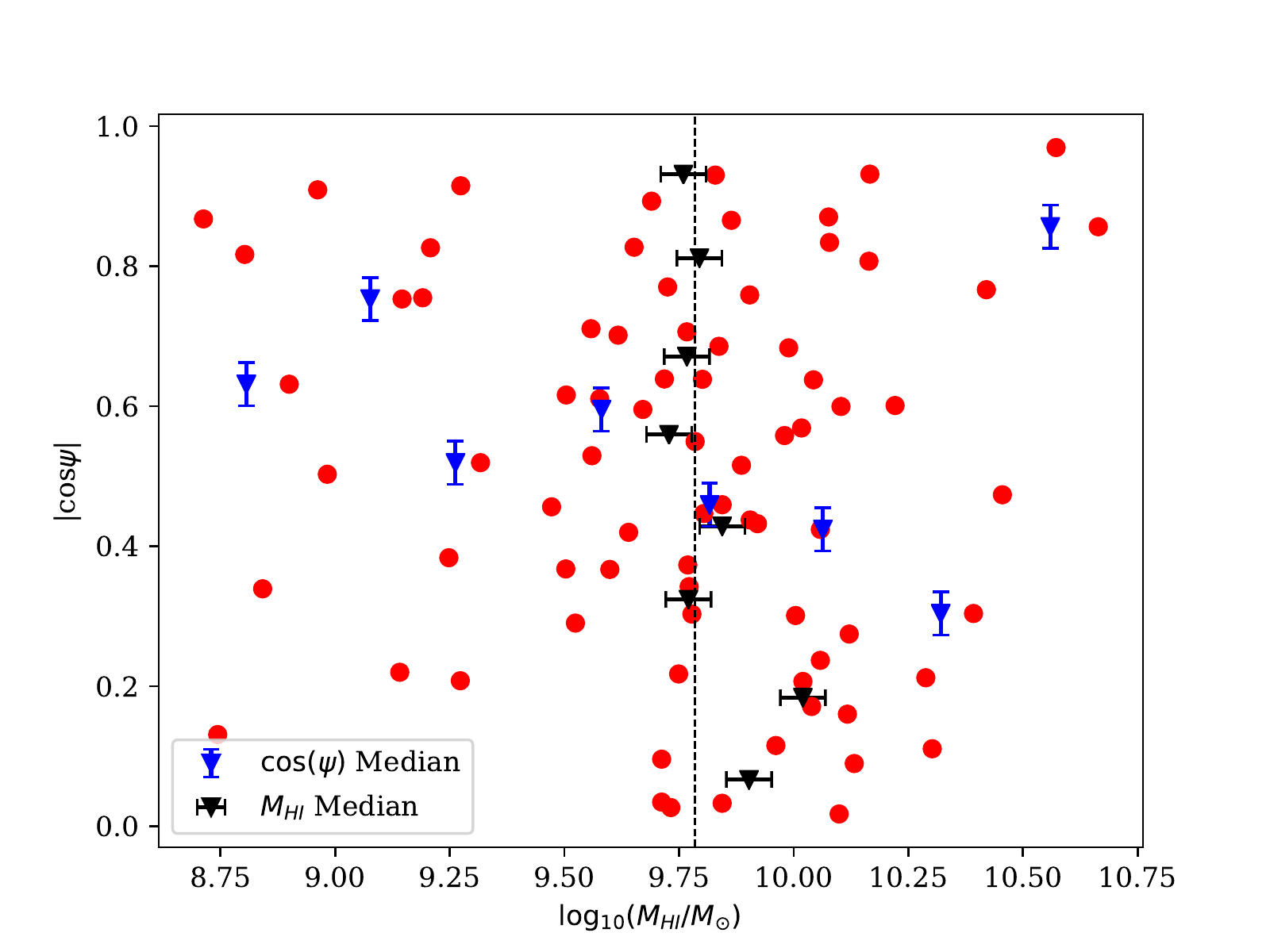}
    \caption{$\lvert \cos \psi \rvert$ for the galaxy sample as a function of H{\sc i} mass computed with mirror boundary conditions. The  vertical black dashed line represents the median of the H{\sc i} mass for our sample. The 77 galaxies were firstly split in 8 bins of $\lvert \cos \psi \rvert$, then they were split in 8 bins of H{\sc i} mass. The black triangles represent the median of the H{\sc i} mass for each bin of $\lvert \cos \psi \rvert$ whilst the blue triangles represent the medians of $\lvert \cos \psi \rvert$ for each bin of $M_{\mathrm{H \sc I}}$.}
    \label{fig:hi-m-mirror}
\end{figure}

We investigate how $\lvert \cos \psi \rvert$ depends on the H{\sc i}-mass of the galaxies. In Figure~\ref{fig:hi-m-mirror}\footnote{We choose not to show the uncertainties on each value of $\cos \psi$ in the remaining figures for clarity, but note that Figure~\ref{fig:hi-p-dist} does show these estimated uncertainties.}, the relationship between $\lvert \cos \psi \rvert$ as a function of H{\sc i} mass is shown. The black vertical dotted line marks the $10^{9.78} M_{\odot}$ value, which represents the median of $M_{\mathrm{H{\sc{I}}}}$ for our sample. This value is close to $10^{9.5} M_{\odot}$, at which the spin transition was observed in the simulation performed by \citet{Kraljic_2020}, which overall agrees with our findings, as the lower left corner of Figure \ref{fig:hi-m-mirror} is less populated compared to the other regions.

However, we find $\langle\lvert \cos \psi \rvert\rangle= 0.52 \pm 0.04$ for galaxies with $\log_{10}(M_{\rm HI}/M_{\odot}) <  9.78$, which is consistent with that of the full sample, where $\langle\lvert \cos \psi \rvert\rangle = 0.51 \pm 0.03$. For the sample with a $\log_{10}(M_{\rm HI}/M_{\odot}) >  9.78$, we find $\langle\lvert \cos \psi \rvert\rangle= 0.50 \pm 0.04$, with the Mann-Whitney U test p-value of 0.4. Therefore, we find no statistical difference in the spin alignment of galaxies with H{\sc i} mass less than or greater than the median value of $10^{9.78}$\,M$_{\odot}$.

Therefore, similar to the study by \citet{bird2019chiles}, we find no evidence for a spin transition at an H{\sc i} mass of $\sim 10^{9.5}$\,M$_{\odot}$. Additionally, we performed a correlation test between H{\sc i} mass and $\lvert \cos \psi \rvert$ to verify the relationship between the parameters. Using Kendall's Tau \citep{ktau} and Spearman Rank correlation \citep{src}, we find no evidence for a correlation: a $\tau$ of $-0.06$ with an associated p-value of $0.5$ and a Spearman Rank coefficient of $-0.08$ with an associated p-value of $0.5$.

\subsection{Spin alignment as a function of other factors}

In addition to distance-to-filament and H{\sc i} mass, we investigate other key properties relating galaxies to their environment: the HI-to-stellar mass ratio and the baryon fraction. We used the ancillary data extracted by the MIGHTEE-HI team for the $ugrizYJHK_s$ photometry as detailed in \cite{Maddox_2021}. The Spectral Energy Distribution (SED) fitting code LePhare \citep{lephare1, lephare2} was then used to derive the stellar mass. The uncertainty in stellar mass for each galaxy that we adopt is $ \sim 0.1$\,dex \citep{adams2021evolution}.

\begin{figure}
	\includegraphics[width=\columnwidth]{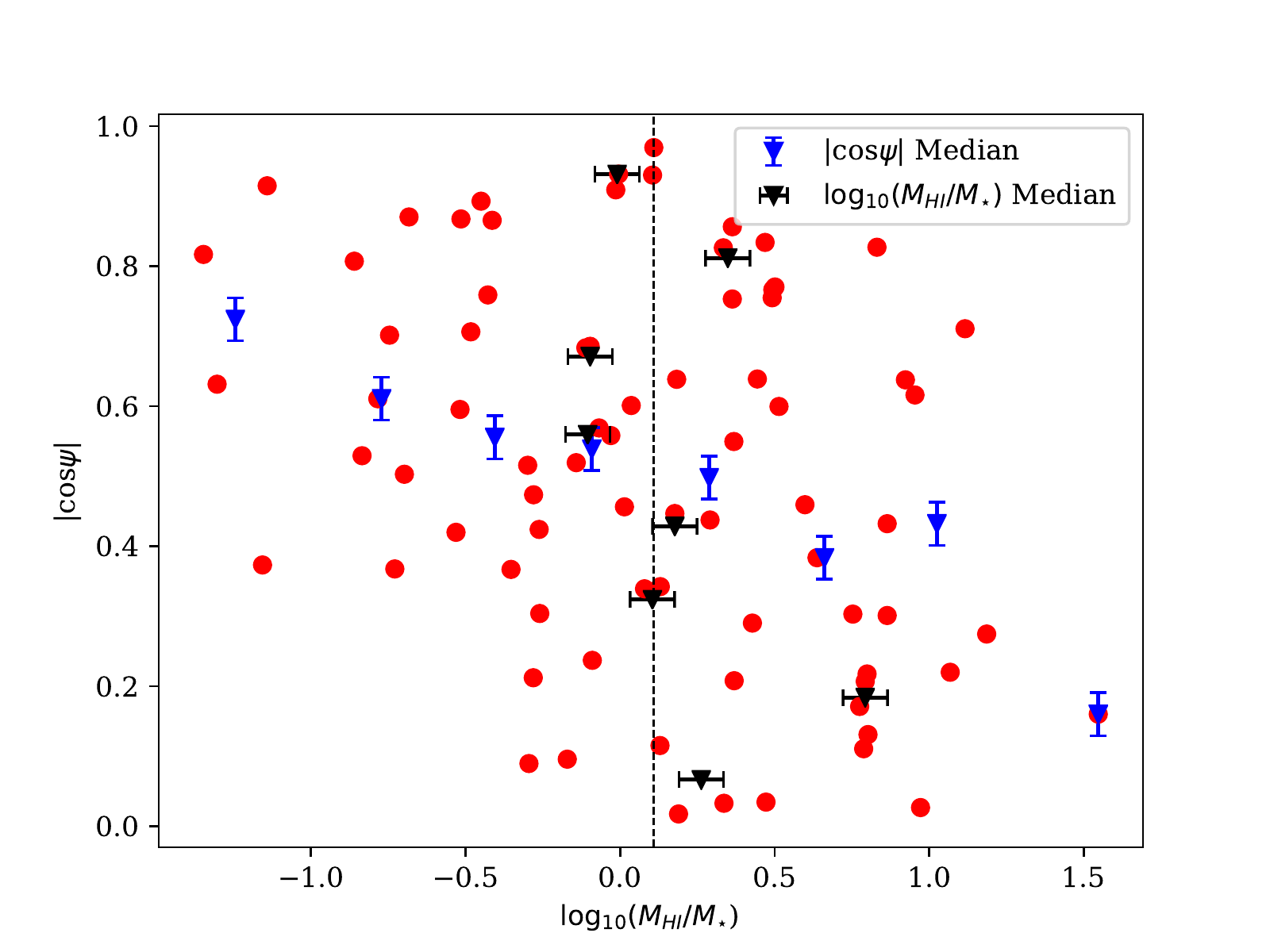}
    \caption{$\lvert \cos \psi \rvert$ for the galaxy sample as a function of log HI-to-stellar mass ratio computed with mirror boundary conditions. The 77 galaxies were firstly split in 8 bins of $\lvert \cos \psi \rvert$, then they were split in 8 bins of $M_{\mathrm{H{\sc I}}}/M_{\star}$. The black triangles represent the medians of log HI-to-stellar mass ratio for each bin whilst the blue triangles represent the medians of $\lvert \cos \psi \rvert$ for each bin. The width of the bins is denoted by the width (height) of the error bars for the blue (black) points.}
    \label{fig:gf-mirror}
\end{figure}

\subsubsection{H{\sc i} -- stellar mass ratio}
The stellar mass of a galaxy is linked to both the environment and the intrinsic properties of the galaxy - for example, galaxies with a higher stellar mass tend to be in dense environments \citep[e.g.][]{vulcani}. 

To determine if there is a difference in the behaviour in the alignment of the spin axis of a galaxy to its nearest filament, we split our sample according to its H{\sc i}-to-stellar mass ratio at the median of the sample ($\log_{10}(M_{\rm HI}/M_{\star})  = 0.11$). For $\log_{10}(M_{\rm HI}/M_{\star})  < 0.11$, we find $\langle\lvert \cos \psi \rvert\rangle= 0.58 \pm 0.04$, which is significantly higher than that for the  sample with $\log_{10}({M_{\rm HI}/M_{\star}})  > 0.11$, $\langle\lvert \cos \psi \rvert\rangle= 0.44 \pm 0.05$ with a Mann-Whitney U test p-value of $0.012$. Figure \ref{fig:gf-mirror} shows the behaviour of the $\lvert \cos \psi \rvert$ as a function of $\log_{10}({M_{\rm HI}/M_{\star}})$ - where we see a dearth of galaxies for $\log_{10}({M_{\rm HI}/M_{\star}})< -0.5$. Thus, we find evidence for the spin changing from aligned to mis-aligned for the galaxies below and above the median of the HI-to-stellar mass ratio $\log_{10}({M_{\rm HI}/M_{\star}}) = 0.11$, respectively. 
Furthermore, the KS test ($p = 7.5 \cdot 10^{-3}$) suggests that for $\log_{10}({M_{\rm HI}/M_{\star}}) < 0.11$, the distribution of $\lvert \cos \psi \rvert$ is not consistent with being drawn from an underlying random galaxy spin alignment distribution.

Both the Kendall Tau and Spearman Rank tests also suggest  a weak correlation between the H{\sc i}-to-stellar mass ratio and the $\lvert \cos \psi \rvert$: a $\tau$ of $-0.209$ with an associated p-value of $0.007$ and a Spearman Rank coefficient of $-0.311$ with an associated p-value of $0.006$.

\begin{figure}
	\includegraphics[width=\columnwidth]{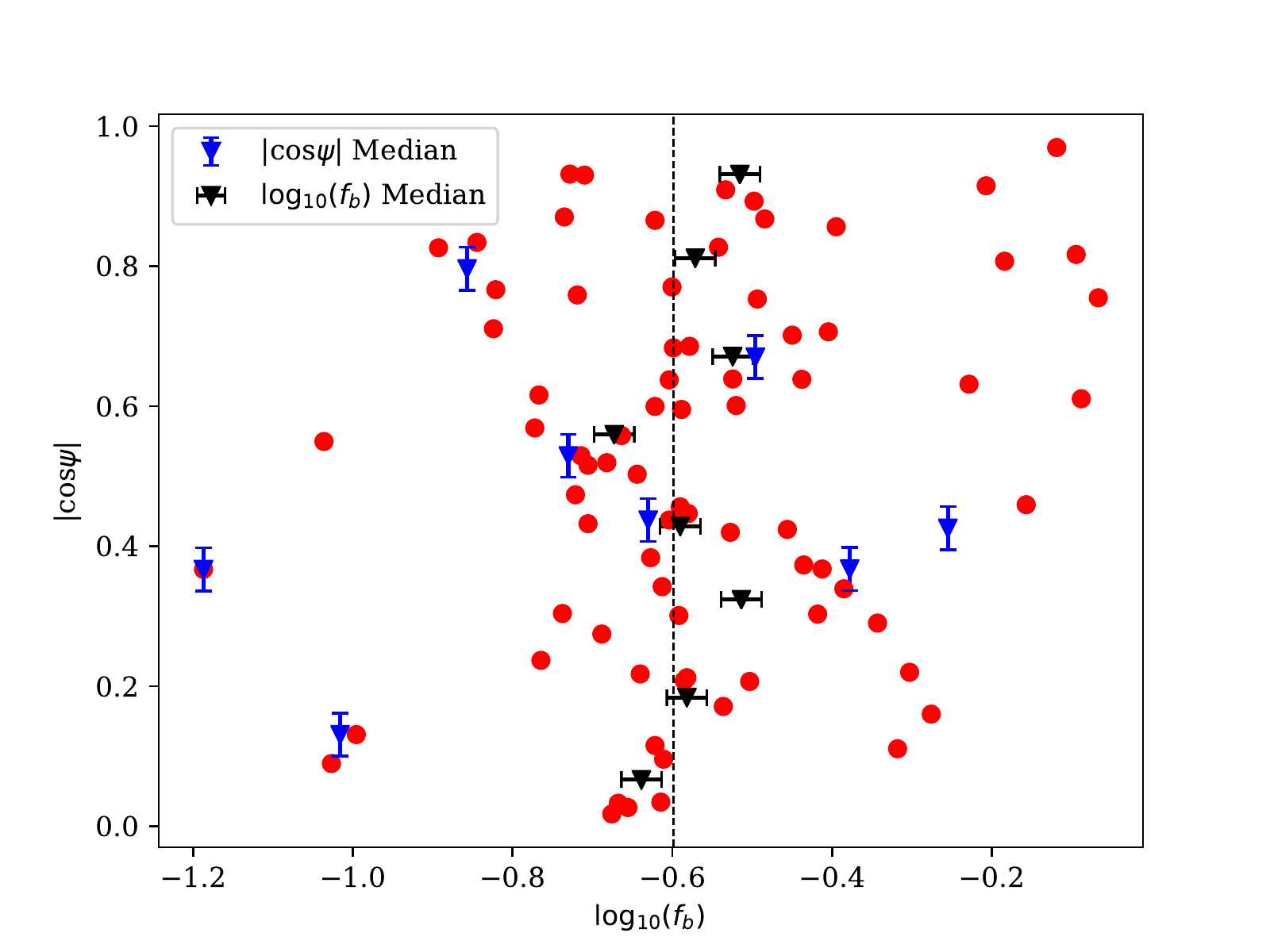}
    \caption{$\lvert \cos \psi \rvert$ as a function of the log of the baryon mass fraction $f_{b} = M_{\mathrm{baryon}}/M_{\mathrm{dyn}}$.  The  vertical black dashed line represents the median $f_b$ for our sample. The 77 galaxies were firstly split in 8 bins of $\lvert \cos \psi \rvert$, then they were split in 8 bins of $f_b$. The black triangles represent the medians of the baryon mass fraction for each bin whilst the blue triangles represent the medians of $\lvert \cos \psi \rvert$ for each bin. The width of the bins is denoted by the width (height) of the error bars for the black (blue) points.}
    \label{fig:fbar-mirror}
\end{figure}

\subsubsection{Baryonic mass versus dynamical mass}

Finally, we explore the effect of the dark matter, through the dynamical mass and the baryon mass fraction. We calculate the dynamical mass $M_{\rm dyn}$ as:
\begin{equation}
    M_{\rm dyn} = \frac{R}{G} V_{\mathrm{rot}}^2,
\end{equation}
where $V_{\mathrm{rot}}$ is the rotational velocity of the galaxy and $R$ is the radius at which the rotational velocity is measured from the resolved H{\sc i} rotation curves, which tend to extend much further than the stellar disk into the dark matter halo \citep[see][for details]{Ponomareva2021}. We then define the baryon mass fraction as $f_b = M_{\mathrm{baryon}}/M_{\rm dyn}$, where $M_{\mathrm{baryon}} = 1.4 \cdot M_{\mathrm{H{\sc{I}}}} + M_{\star}$. The factor of 1.4 is included to account for the primordial abundance of metals and helium \citep{arnett}, however it does not include the molecular gas component.

Following the same method as above, we split the galaxy sample at the median of the baryon mass fraction, $\log_{10}(f_b) = -0.6$. Here, the average value of $\lvert \cos \psi \rvert$ tends to be marginally mis-aligned for galaxies with a lower baryon mass fraction: $\langle\lvert \cos \psi \rvert\rangle= 0.47 \pm 0.05$, whilst for the higher baryon mass fraction, it tends towards alignment: $\langle\lvert \cos \psi \rvert\rangle= 0.55 \pm 0.04$. Figure \ref{fig:fbar-mirror} shows $\lvert \cos \psi \rvert$ as a function of the baryon mass fraction - it can be seen that for the lower baryon mass fraction there is a spread in $ \cos \psi$, whilst for the highest baryon mass fraction ($\log_{10}(f_b) > -0.2$), the galaxies all tend to be aligned. This implies that galaxies with high stellar and/or H{\sc i} mass tend to retain their alignment with the filament, whilst the galaxies with a higher dark matter fraction are less likely to be aligned with the filaments. However, we find no evidence for a correlation using both the Kendall's Tau $\tau = 0.07$ and the Spearman Rank coefficient ($0.11$), with associated p-values of $0.38$ and $0.36$, respectively. Similarly to the $M_{\mathrm{H{\sc{I}}}}/M_{\star}$ , we notice a spin transition at the median value of the baryon mass fraction $\log_{10}(f_b) = -0.6$, from $\langle\lvert \cos \psi \rvert\rangle= 0.47 \pm 0.05$ to $\langle\lvert \cos \psi \rvert\rangle= 0.55 \pm 0.04$, although the Mann-Whitney U p-value of $0.132$ shows that this is not significant given the current sample. Therefore, galaxies with a lower dark matter content are more likely to be aligned, but for the rest of the galaxies the alignment is consistent with being random. This will require more data and future analysis to see if any trend exists.

\begin{table}
\centering
\caption{The mean of the cosine of the angle between the H{\sc i} spin of the galaxy and its closest filament $\cos{\psi}$ and comparison parameters: H{\sc i} Mass $M_{\mathrm{H{\sc{I}}}}$, H{\sc i}-to-stellar mass ratio $M_{\mathrm{H{\sc{I}}}}/M_{\star}$ and baryon mass fraction $f_b$.}
\label{table:galaxy-prop-m}
\begin{tabular}{cccc}
\hline\hline
\textbf{Parameter}       & \textbf{Cut} & ${\langle\lvert\mathbf{cos\psi}\rvert\rangle} $ &  $\mathbf{p_{MW}}$\\  \hline\hline
\multirow{2}{*}{$\log_{10}\Big(\frac{M_{\mathrm{H{\sc{I}}}}}{M_{\odot}}\Big)$} &  $ < 9.78$ &  $0.52 \pm 0.04$  & \multirow{2}{*}{$0.40$}\\
                            & $ > 9.78$   &  $0.50 \pm 0.05$ & \\  \hline
\multirow{2}{*}{$\log_{10}\Big(\frac{M_{\mathrm{H{\sc{I}}}}}{M_{\star}}\Big)$} & $ < 0.11$ & $0.58 \pm 0.04$&\multirow{2}{*}{$0.01$}\\
                            & $ > 0.11$   &  $0.44 \pm 0.05$ & \\ \hline
\multirow{2}{*}{$\log_{10}(f_b)$}&  $ < -0.598$ &  $0.47 \pm 0.05$  & \multirow{2}{*}{$0.13$} \\
                        & $ > -0.598$   &  $0.55 \pm 0.04$ &  \\
\hline
\end{tabular}
\end{table}

\begin{figure}
  \centering
  \captionsetup{justification=centering}
  \includegraphics[width=1\linewidth]{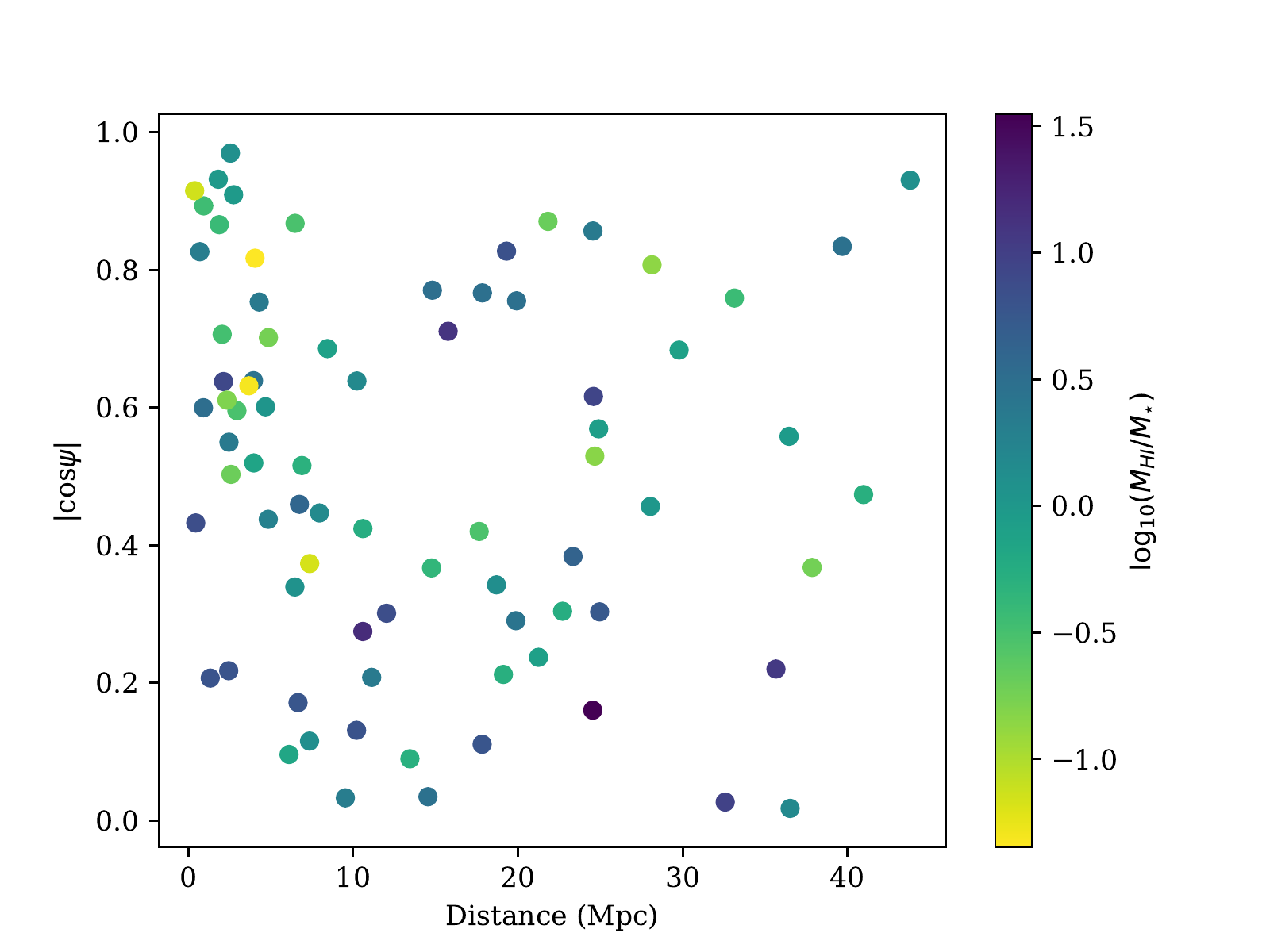}  
\caption{$\lvert \cos \psi \rvert$ as a function of distance as presented in Figure \ref{fig:hi-dist}, with the H{\sc i}-stellar mass ratio $\log_{10}(M_{\mathrm{H{\sc{I}}}}/M_{\star})$ overlaid as a colourmap.}
  \label{fig:cos-dist-gf}
\end{figure}

\subsection{Discussion}
\begin{figure}
	\includegraphics[width=\columnwidth]{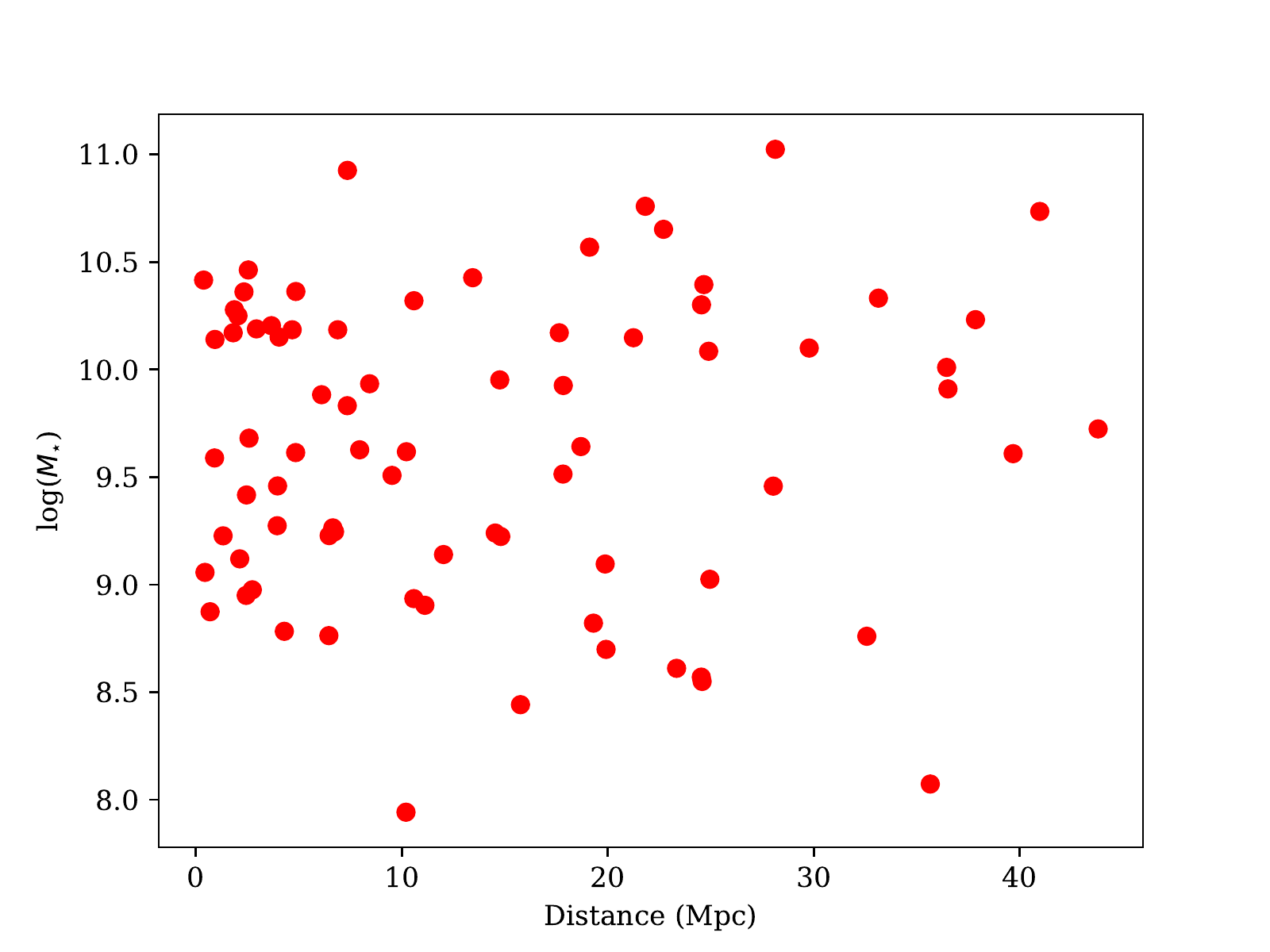}
    \caption{The stellar mass for the galaxy sample as a function of distance to the closest filament computed with mirror boundary conditions.}
    \label{fig:mstar-mirror}
\end{figure}

Taken together, our results suggest that the stellar mass of a galaxy has a strong influence on the spin of the galaxy in relation to the filaments. In Figure~\ref{fig:cos-dist-gf} we show the $M_{\rm HI} / M_{\star}$ ratio as a function of the distance from the nearest filament. It shows that those galaxies with the lowest $M_{\rm HI} / M_{\star}$ ratio, and therefore the highest stellar mass, given that our sample is selected on H{\sc i} mass, tend to be aligned. The stellar mass being an important influence on the spin would be consistent with several simulations \citep{dubois, Kraljic_2020}, where they find a transition at a stellar mass of $M_{\star} \sim 10^{10}$\,M$_{\odot}$, from aligned to mis-aligned. Furthermore, \citet{welker} found a similar spin transition between lower mass and higher mass galaxies using the SAMI survey around $10^{10.4} M_{\odot} - 10^{10.9} M_{\odot}$. To understand whether there is a bias, such that galaxies with higher stellar mass tend to be found in denser environments and closer to the filaments, we checked their position with respect to their closest filament. As can be seen in Figure \ref{fig:mstar-mirror}, the stellar mass of the galaxies is randomly distributed, therefore we find no evidence that this could be the reason for our results on the spin-alignment between galaxies and their closest filaments. Due to the fact that our sample is H{\sc i} selected, comparisons with other studies using stellar mass would be biased and uninformative for our study.

We also investigate whether the dark matter content affects the relation between the spin-axis of the galaxy and the orientation of its nearest filament by determining the baryonic mass fraction. We find that for the highest baryonic mass fractions that the galaxies tend to align with the nearest filament. However, we find no evidence of a trend of alignment with baryonic mass fraction. Both of these reinforce the evidence from the $M_{\rm HI} / M_{\star}$ ratio, that the stellar mass is a key factor in determining whether a galaxy is aligned with its nearest filament or not.

We also note that some of the trends that we do find could also be linked with the morphology of the galaxies - as the baryon mass fraction varies depending whether a galaxy is elliptical or spiral. As mentioned before, \citet{Kraljic_2021} found a dependence of the shape of the galaxy and its alignment with the filament - S0 type galaxies are more likely to be mis-aligned. Other environmental factors, such as the gas inflows around the filaments or galaxy mergers in the filament regions could also influence the alignment of the spin of the galaxies. \citet{welker_2014} have shown that galaxies that have undergone fewer mergers are more likely to be aligned with the filaments, whilst galaxies that have undergone mergers along the filaments are more likely to have their spins swung to mis-alignment - especially for major mergers. Given that it is more likely for the more massive galaxies to have undergone a merger in the past, the fact that in our H{\sc i}-selected sample it is those galaxies with a relatively larger stellar mass which tend to be more aligned, appears to be at odds with this finding from simulations. However, it is difficult to interpret our results in this context due to the fact that our sample is H{\sc i}-selected and dominated by relatively low-stellar mass objects and contains very few objects with masses above the mass where the spin transition appears in observations and simulations \citep{welker}. 
However, we also expect gas-rich mergers to increase the amount of neutral gas in galaxies \citep{Ellison2018} and this may provide an explanation of our results, where we see that those galaxies which are misaligned do have higher H{\sc i} mass fractions compared to their stellar mass. Such a merger history may also be apparent in the stellar populations of the galaxies and thus a fruitful future line of enquiry would be to investigate whether the aligned and misaligned populations exhibit different star formation histories, ongoing enhanced star formation or morphological evidence of a merger event happening.

It is difficult to explore this with the current sample due to the limited number of objects, as we do not have the number statistics, or the filament constraints. However, it will be possible as the MIGHTEE survey expands to the full 20 deg$^{2}$ area, substantially increasing the sample size we would have to work with. For the $5$\,deg$^2$ in the COSMOS $+$ XMM-LSS Early Science data, which is not to full depth and was taken with a coarser channel width than will be done for the rest of the survey, we have about $50$ detections per square degree. For the full MIGHTEE survey, we expect around $\sim 1000$ detections at $z < 0.1$ \citep{Maddox_2021}, as such we will be able to investigate sub-samples of the galaxies in relation to their filaments, binning with respect to morphology, age and also the actual spin and angular momentum of the galaxies within the sample.

\begin{table}
\centering
\caption{The coefficients and p-values for the two correlation tests, Kendall's Tau and Spearman Rank, for each parameter against the $\lvert \cos \psi \rvert$.}
\label{table:galaxy-prop-stats-m}
\begin{tabular}{ccccc}
\hline\hline
 \multirow{2}{*}{\textbf{Parameter}  }                             & \multicolumn{2}{c}{\textbf{Kendall's Tau}} & \multicolumn{2}{c}{\textbf{Spearman Rank}} \\ 
                                                                        & \textbf{$\tau$}   & \textbf{p-value}  & \textbf{coefficient}   & \textbf{p-value}  \\ \hline\hline

   Distance & $-0.144$ & $0.065$ & $-0.205$ & $0.074$\\
   $M_{\mathrm{H{\sc{I}}}}$ & $-0.058$ & $0.452$ & $-0.083$ & $0.472$\\
   $M_{\mathrm{H{\sc{I}}}}/M_{\star}$ & $-0.209$ & $0.007$ & $-0.311$ & $0.006$\\
   $f_b$ & $0.069$ & $0.377$ & $0.107$ & $0.355$\\ \hline
\end{tabular}
\end{table}
\section{Conclusions}
\label{sec:conclusions}

In this work, we present a study of the 3D spins of 77 H{\sc i} galaxies identified with the MIGHTEE-HI survey, and their link to the filaments of the cosmic web. The large-scale filaments are computed using DisPerSE and optical galaxies from the COSMOS and XMM-LSS fields. We took into consideration several parameters that might affect the alignment between the galaxies and the filaments and found that:
\begin{itemize}
    \item distance-to-filament: galaxies closer ($<5$\,Mpc) to the spine of the filament tend to be aligned with their nearest filament.
    \item H{\sc i} content of galaxy: no spin transition was found using the H{\sc i} mass of the galaxy for mirror boundary conditions for $\log_{10}(M_{\mathrm{H{\sc{I}}}}/M_{\odot}) < 9.78$.
    \item H{\sc i}-to-stellar mass ratio of galaxy: we find a preference for alignment for the galaxies with a lower H{\sc i}-to-stellar mass ratio and overall throughout the sample, as well as a spin transition.
    \item baryon mass fraction of galaxy: we find those galaxies with the highest baryon mass fraction to exhibit alignment with their nearest filament. However, we find no trend across the range of baryonic mass fraction probed. 
\end{itemize}

Overall, we find the compelling evidence that the neutral gas fraction relative to the stellar mass of a galaxy is clearly related to the alignment of the galaxy spin vector and the nearest filament. Furthermore, galaxies show greater evidence for alignment the closer they are to the filament, suggesting that there is an interplay between the galaxy spin axis and the filament. 
We suggest that this is due to those galaxies which have undergone a recent gas-rich merger have their spin-orientation disrupted with respect to the filament, whereas those galaxies which have not undergone a recent merger tend to retain their alignment and their evolution is dictated by secular processes. Such a scenario could be investigated further by measuring the star-formation histories of the galaxies as a function of their spin alignment with the filaments.
Given that the number statistics in this study are limited, it would benefit from additional data. However, it underlines the potential of the MIGHTEE Large Survey Program, as well as the MeerKAT telescope. With more data expected in the coming years , the sample size of H{\sc i} galaxies will increase significantly and enable a big step forward in understanding how galaxies are powered by fuel drawn from the cosmic web. Whilst a study with respect to redshift will be difficult with MIGHTEE alone, combining information from MIGHTEE with the deeper and narrower Looking At the Distant Universe with the MeerKAT Array \citep[LADUMA;][]{Blyth2016} may provide the necessary redshift baseline. However, the need to at least marginally resolve galaxies for the kinematic modelling would limit the sample to the largest or most H{\sc i}-rich galaxies, given MeerKAT's synthesised beam.

\section*{Acknowledgements}

We thank the anonymous referee for their useful comments and feedback. We thank Thijs van der Hulst for insightful comments. MNT, MJJ and IW acknowledge support from the Oxford Hintze Centre for Astrophysical Surveys which is funded through generous support from the Hintze Family Charitable Foundation. IH, MJJ and AAP acknowledge support from the UK Science and Technology Facilities Council [ST/N000919/1]. IH acknowledges support from the South African Radio Astronomy Observatory which is a facility of the National Research Foundation (NRF), an agency of the Department of Science and Innovation. NM acknowledges support of the LMU Faculty of Physics. The MeerKAT telescope is operated by the South African Radio Astronomy Observatory, which is a facility of the National Research Foundation, an agency of the Department of Science and Innovation. We acknowledge use of the Inter-University Institute for Data Intensive Astronomy (IDIA) data intensive research cloud for data processing. IDIA is a South African university partnership involving the University of Cape Town, the University of Pretoria and the University of the Western Cape. The authors acknowledge the Centre for High Performance Computing (CHPC), South Africa, for providing computational resources to this research project. This work is based on data products from observations made with ESO Telescopes at the La Silla Paranal Observatory under ESO programme ID 179.A-2005 (Ultra-VISTA) and ID 179.A- 2006(VIDEO) and on data products produced by CALET and the Cambridge Astronomy Survey Unit on behalf of the Ultra-VISTA and VIDEO consortia. Based on observations obtained with MegaPrime/MegaCam, a joint project of CFHT and CEA/IRFU, at the Canada-France-Hawaii Telescope (CFHT) which is operated by the National Research Council (NRC) of Canada, the Institut National des Science de l’Univers of the Centre National de la Recherche Scientifique (CNRS) of France, and the University of Hawaii. This work is based in part on data products produced at Terapix available at the Canadian Astronomy Data Centre as part of the Canada-France-Hawaii Telescope Legacy Survey, a collaborative project of NRC and CNRS. The Hyper Suprime-Cam (HSC) collaboration includes the astronomical communities of Japan and Taiwan, and Princeton University. The HSC instrumentation and soft- ware were developed by the National Astronomical Observatory of Japan (NAOJ), the Kavli Institute for the Physics and Mathematics of the Universe (Kavli IPMU), the University of Tokyo, the High Energy Accelerator Research Organization (KEK), the Academia Sinica Institute for Astronomy and Astrophysics in Taiwan (ASIAA), and Princeton University. Funding was contributed by the FIRST program from Japanese Cabinet Office, the Ministry of Education, Culture, Sports, Science and Technology (MEXT), the Japan Society for the Promotion of Science (JSPS), Japan Science and Technology Agency (JST), the Toray Science Foundation, NAOJ, Kavli IPMU, KEK, ASIAA, and Princeton University. 

This research made use of Astropy,\footnote{\url{http://www.astropy.org}} a community-developed core Python package for Astronomy \citep{astropy:2013, astropy:2018}.

\section*{Data Availability}
The MIGHTEE-HI spectral cubes will be released as part of the first data release of the MIGHTEE survey, which will include cubelets of the sources discussed in this paper. The derived quantities from the multi-wavelength ancillary data was released with the final data release of the VIDEO survey mid 2021. Alternative products are already available from the Herschel Extragalactic Legacy Project \citep[HELP;][]{shirley2021help} and also soon from the Deep Extragalactic VIsible Legacy Survey \citep[DEVILS;][]{devils}.



\bibliographystyle{mnras}
\bibliography{ref} 

\begin{thebibliography}{}
\makeatletter
\relax
\def\mn@urlcharsother{\let\do\@makeother \do\$\do\&\do\#\do\^\do\_\do\%\do\~}
\def\mn@doi{\begingroup\mn@urlcharsother \@ifnextchar [ {\mn@doi@}
  {\mn@doi@[]}}
\def\mn@doi@[#1]#2{\def\@tempa{#1}\ifx\@tempa\@empty \href
  {http://dx.doi.org/#2} {doi:#2}\else \href {http://dx.doi.org/#2} {#1}\fi
  \endgroup}
\def\mn@eprint#1#2{\mn@eprint@#1:#2::\@nil}
\def\mn@eprint@arXiv#1{\href {http://arxiv.org/abs/#1} {{\tt arXiv:#1}}}
\def\mn@eprint@dblp#1{\href {http://dblp.uni-trier.de/rec/bibtex/#1.xml}
  {dblp:#1}}
\def\mn@eprint@#1:#2:#3:#4\@nil{\def\@tempa {#1}\def\@tempb {#2}\def\@tempc
  {#3}\ifx \@tempc \@empty \let \@tempc \@tempb \let \@tempb \@tempa \fi \ifx
  \@tempb \@empty \def\@tempb {arXiv}\fi \@ifundefined
  {mn@eprint@\@tempb}{\@tempb:\@tempc}{\expandafter \expandafter \csname
  mn@eprint@\@tempb\endcsname \expandafter{\@tempc}}}

\bibitem[\protect\citeauthoryear{Adams, Bowler, Jarvis, Häußler  \&
  Lagos}{Adams et~al.}{2021}]{adams2021evolution}
Adams N.~J.,  Bowler R. A.~A.,  Jarvis M.~J.,  Häußler B.,   Lagos C. D.~P.,
  2021, \mn@doi [Monthly Notices of the Royal Astronomical Society]
  {10.1093/mnras/stab1956}, 506, 4933–4951

\bibitem[\protect\citeauthoryear{{Aihara} et~al.,}{{Aihara}
  et~al.}{2018}]{Aihara2018}
{Aihara} H.,  et~al., 2018, \mn@doi [\pasj] {10.1093/pasj/psx081}, \href
  {https://ui.adsabs.harvard.edu/abs/2018PASJ...70S...8A} {70, S8}

\bibitem[\protect\citeauthoryear{{Arag{\'o}n-Calvo}, {van de Weygaert}, {Jones}
   \& {van der Hulst}}{{Arag{\'o}n-Calvo} et~al.}{2007}]{aragon-calvo_2007}
{Arag{\'o}n-Calvo} M.~A.,  {van de Weygaert} R.,  {Jones} B. J.~T.,   {van der
  Hulst} J.~M.,  2007, \mn@doi [\apjl] {10.1086/511633}, \href
  {https://ui.adsabs.harvard.edu/abs/2007ApJ...655L...5A} {655, L5}

\bibitem[\protect\citeauthoryear{{Arnett}}{{Arnett}}{1999}]{arnett}
{Arnett} D.,  1999, \mn@doi [\apss] {10.1023/A:1002131915162}, \href
  {https://ui.adsabs.harvard.edu/abs/1999Ap&SS.265...29A} {265, 29}

\bibitem[\protect\citeauthoryear{{Arnouts}, {Cristiani}, {Moscardini},
  {Matarrese}, {Lucchin}, {Fontana}  \& {Giallongo}}{{Arnouts}
  et~al.}{1999}]{lephare1}
{Arnouts} S.,  {Cristiani} S.,  {Moscardini} L.,  {Matarrese} S.,  {Lucchin}
  F.,  {Fontana} A.,   {Giallongo} E.,  1999, \mn@doi [\mnras]
  {10.1046/j.1365-8711.1999.02978.x}, \href
  {https://ui.adsabs.harvard.edu/abs/1999MNRAS.310..540A} {310, 540}

\bibitem[\protect\citeauthoryear{{Astropy Collaboration} et~al.,}{{Astropy
  Collaboration} et~al.}{2013}]{astropy:2013}
{Astropy Collaboration} et~al., 2013, \mn@doi [\aap]
  {10.1051/0004-6361/201322068}, \href
  {http://adsabs.harvard.edu/abs/2013A%26A...558A..33A} {558, A33}

\bibitem[\protect\citeauthoryear{{Astropy Collaboration} et~al.,}{{Astropy
  Collaboration} et~al.}{2018}]{astropy:2018}
{Astropy Collaboration} et~al., 2018, \mn@doi [\aj] {10.3847/1538-3881/aabc4f},
  \href {https://ui.adsabs.harvard.edu/abs/2018AJ....156..123A} {156, 123}

\bibitem[\protect\citeauthoryear{{Blue Bird} et~al.,}{{Blue Bird}
  et~al.}{2020}]{bird2019chiles}
{Blue Bird} J.,  et~al., 2020, Monthly Notices of the Royal Astronomical
  Society, 492, 153

\bibitem[\protect\citeauthoryear{{Blyth} et~al.,}{{Blyth}
  et~al.}{2016}]{Blyth2016}
{Blyth} S.,  et~al., 2016, in MeerKAT Science: On the Pathway to the SKA. p.~4

\bibitem[\protect\citeauthoryear{{Bond}, {Kofman}  \& {Pogosyan}}{{Bond}
  et~al.}{1996}]{bond-1996}
{Bond} J.~R.,  {Kofman} L.,   {Pogosyan} D.,  1996, \mn@doi [Nature
  Astrophysics] {10.1038/380603a0}, \href
  {https://ui.adsabs.harvard.edu/abs/1996Natur.380..603B} {380, 603}

\bibitem[\protect\citeauthoryear{{Bundy} et~al.,}{{Bundy} et~al.}{2015}]{manga}
{Bundy} K.,  et~al., 2015, \mn@doi [\apj] {10.1088/0004-637X/798/1/7}, \href
  {https://ui.adsabs.harvard.edu/abs/2015ApJ...798....7B} {798, 7}

\bibitem[\protect\citeauthoryear{{Codis}, {Pichon}  \& {Pogosyan}}{{Codis}
  et~al.}{2015}]{codis}
{Codis} S.,  {Pichon} C.,   {Pogosyan} D.,  2015, \mn@doi [\mnras]
  {10.1093/mnras/stv1570}, \href
  {https://ui.adsabs.harvard.edu/abs/2015MNRAS.452.3369C} {452, 3369}

\bibitem[\protect\citeauthoryear{Colless et~al.,}{Colless
  et~al.}{2001}]{2dfgrs}
Colless M.,  et~al., 2001, \mn@doi [Monthly Notices of the Royal Astronomical
  Society] {10.1046/j.1365-8711.2001.04902.x}, 328, 1039–1063

\bibitem[\protect\citeauthoryear{{Crone Odekon}, {Hallenbeck}, {Haynes},
  {Koopmann}, {Phi}  \& {Wolfe}}{{Crone Odekon} et~al.}{2018}]{crone-odekon}
{Crone Odekon} M.,  {Hallenbeck} G.,  {Haynes} M.~P.,  {Koopmann} R.~A.,  {Phi}
  A.,   {Wolfe} P.-F.,  2018, \mn@doi [\apj] {10.3847/1538-4357/aaa1e8}, \href
  {https://ui.adsabs.harvard.edu/abs/2018ApJ...852..142C} {852, 142}

\bibitem[\protect\citeauthoryear{Croom et~al.,}{Croom et~al.}{2012}]{sami}
Croom S.~M.,  et~al., 2012, \mn@doi [Monthly Notices of the Royal Astronomical
  Society] {10.1111/j.1365-2966.2011.20365.x}

\bibitem[\protect\citeauthoryear{{Dav{\'e}}, {Angl{\'e}s-Alc{\'a}zar},
  {Narayanan}, {Li}, {Rafieferantsoa}  \& {Appleby}}{{Dav{\'e}}
  et~al.}{2019a}]{simba-sim}
{Dav{\'e}} R.,  {Angl{\'e}s-Alc{\'a}zar} D.,  {Narayanan} D.,  {Li} Q.,
  {Rafieferantsoa} M.~H.,   {Appleby} S.,  2019a, \mn@doi [\mnras]
  {10.1093/mnras/stz937}, \href
  {https://ui.adsabs.harvard.edu/abs/2019MNRAS.486.2827D} {486, 2827}

\bibitem[\protect\citeauthoryear{{Dav{\'e}}, {Angl{\'e}s-Alc{\'a}zar},
  {Narayanan}, {Li}, {Rafieferantsoa}  \& {Appleby}}{{Dav{\'e}}
  et~al.}{2019b}]{simba}
{Dav{\'e}} R.,  {Angl{\'e}s-Alc{\'a}zar} D.,  {Narayanan} D.,  {Li} Q.,
  {Rafieferantsoa} M.~H.,   {Appleby} S.,  2019b, \mn@doi [Monthly Notices of
  the Royal Astronomical Society] {10.1093/mnras/stz937}, \href
  {https://ui.adsabs.harvard.edu/abs/2019MNRAS.486.2827D} {486, 2827}

\bibitem[\protect\citeauthoryear{{Davies} et~al.,}{{Davies}
  et~al.}{2021}]{devils}
{Davies} L.~J.~M.,  et~al., 2021, \mn@doi [\mnras] {10.1093/mnras/stab1601},
  \href {https://ui.adsabs.harvard.edu/abs/2021MNRAS.506..256D} {506, 256}

\bibitem[\protect\citeauthoryear{{Davis}, {Huchra}, {Latham}  \&
  {Tonry}}{{Davis} et~al.}{1982}]{Davis1982}
{Davis} M.,  {Huchra} J.,  {Latham} D.~W.,   {Tonry} J.,  1982, \mn@doi [\apj]
  {10.1086/159646}, \href
  {https://ui.adsabs.harvard.edu/abs/1982ApJ...253..423D} {253, 423}

\bibitem[\protect\citeauthoryear{Driver et~al.,}{Driver et~al.}{2009}]{gama}
Driver S.~P.,  et~al., 2009, \mn@doi [Astronomy & Geophysics]
  {10.1111/j.1468-4004.2009.50512.x}, 50, 5.12

\bibitem[\protect\citeauthoryear{{Dubois} et~al.,}{{Dubois}
  et~al.}{2014a}]{dubois}
{Dubois} Y.,  et~al., 2014a, \mn@doi [\mnras] {10.1093/mnras/stu1227}, \href
  {https://ui.adsabs.harvard.edu/abs/2014MNRAS.444.1453D} {444, 1453}

\bibitem[\protect\citeauthoryear{{Dubois} et~al.,}{{Dubois}
  et~al.}{2014b}]{horizon-agn}
{Dubois} Y.,  et~al., 2014b, \mn@doi [Monthly Notices of the Royal Astronomical
  Society] {10.1093/mnras/stu1227}, \href
  {https://ui.adsabs.harvard.edu/abs/2014MNRAS.444.1453D} {444, 1453}

\bibitem[\protect\citeauthoryear{{Ellison}, {Catinella}  \&
  {Cortese}}{{Ellison} et~al.}{2018}]{Ellison2018}
{Ellison} S.~L.,  {Catinella} B.,   {Cortese} L.,  2018, \mn@doi [\mnras]
  {10.1093/mnras/sty1247}, \href
  {https://ui.adsabs.harvard.edu/abs/2018MNRAS.478.3447E} {478, 3447}

\bibitem[\protect\citeauthoryear{Fern{\'{a}}ndez et~al.,}{Fern{\'{a}}ndez
  et~al.}{2016}]{chiles}
Fern{\'{a}}ndez X.,  et~al., 2016, \mn@doi [The Astrophysical Journal]
  {10.3847/2041-8205/824/1/l1}, 824, L1

\bibitem[\protect\citeauthoryear{Galárraga-Espinosa, Aghanim, Langer, Gouin
  \& Malavasi}{Galárraga-Espinosa et~al.}{2020}]{Galarraga_Espinosa_2020}
Galárraga-Espinosa D.,  Aghanim N.,  Langer M.,  Gouin C.,   Malavasi N.,
  2020, \mn@doi [Astronomy \& Astrophysics] {10.1051/0004-6361/202037986}, 641,
  A173

\bibitem[\protect\citeauthoryear{{Ganeshaiah Veena}, {Cautun}, {van de
  Weygaert}, {Tempel}, {Jones}, {Rieder}  \& {Frenk}}{{Ganeshaiah Veena}
  et~al.}{2018}]{ganeshaiah-veena-2018}
{Ganeshaiah Veena} P.,  {Cautun} M.,  {van de Weygaert} R.,  {Tempel} E.,
  {Jones} B. J.~T.,  {Rieder} S.,   {Frenk} C.~S.,  2018, \mn@doi [\mnras]
  {10.1093/mnras/sty2270}, \href
  {https://ui.adsabs.harvard.edu/abs/2018MNRAS.481..414G} {481, 414}

\bibitem[\protect\citeauthoryear{Glasser \& Winter}{Glasser \&
  Winter}{1961}]{src}
Glasser G.~J.,  Winter R.~F.,  1961, Biometrika, 48, 444

\bibitem[\protect\citeauthoryear{{Hahn}, {Teyssier}  \& {Carollo}}{{Hahn}
  et~al.}{2010}]{hahn}
{Hahn} O.,  {Teyssier} R.,   {Carollo} C.~M.,  2010, \mn@doi [\mnras]
  {10.1111/j.1365-2966.2010.16494.x}, \href
  {https://ui.adsabs.harvard.edu/abs/2010MNRAS.405..274H} {405, 274}

\bibitem[\protect\citeauthoryear{Heywood et~al.,}{Heywood
  et~al.}{2021}]{Heywood2021}
Heywood I.,  et~al., 2021, \mn@doi [Monthly Notices of the Royal Astronomical
  Society] {10.1093/mnras/stab3021}, 509, 2150

\bibitem[\protect\citeauthoryear{{Hirv, A.}, {Pelt, J.}, {Saar, E.}, {Tago,
  E.}, {Tamm, A.}, {Tempel, E.}  \& {Einasto, M.}}{{Hirv, A.}
  et~al.}{2017}]{hirv}
{Hirv, A.} {Pelt, J.} {Saar, E.} {Tago, E.} {Tamm, A.} {Tempel, E.}  {Einasto,
  M.} 2017, \mn@doi [A\&A] {10.1051/0004-6361/201629248}, 599, A31

\bibitem[\protect\citeauthoryear{{Huchra} et~al.,}{{Huchra}
  et~al.}{2012}]{2mass}
{Huchra} J.~P.,  et~al., 2012, \mn@doi [\apjs] {10.1088/0067-0049/199/2/26},
  \href {https://ui.adsabs.harvard.edu/abs/2012ApJS..199...26H} {199, 26}

\bibitem[\protect\citeauthoryear{{Ilbert} et~al.,}{{Ilbert}
  et~al.}{2006}]{lephare2}
{Ilbert} O.,  et~al., 2006, \mn@doi [\aap] {10.1051/0004-6361:20065138}, \href
  {https://ui.adsabs.harvard.edu/abs/2006A&A...457..841I} {457, 841}

\bibitem[\protect\citeauthoryear{{Jarvis} et~al.,}{{Jarvis}
  et~al.}{2013}]{Jarvis2013}
{Jarvis} M.~J.,  et~al., 2013, \mn@doi [\mnras] {10.1093/mnras/sts118}, \href
  {https://ui.adsabs.harvard.edu/abs/2013MNRAS.428.1281J} {428, 1281}

\bibitem[\protect\citeauthoryear{{Jarvis} et~al.,}{{Jarvis}
  et~al.}{2016}]{mightee}
{Jarvis} M.,  et~al., 2016, in MeerKAT Science: On the Pathway to the SKA. p.~6
  (\mn@eprint {arXiv} {1709.01901})

\bibitem[\protect\citeauthoryear{Jonas}{Jonas}{2009}]{meerkat}
Jonas J.~L.,  2009, \mn@doi [Proceedings of the IEEE]
  {10.1109/JPROC.2009.2020713}, 97, 1522

\bibitem[\protect\citeauthoryear{Kendall}{Kendall}{1938}]{ktau}
Kendall M.~G.,  1938, \mn@doi [Biometrika] {10.1093/biomet/30.1-2.81}, 30, 81

\bibitem[\protect\citeauthoryear{{Kleiner}, {Pimbblet}, {Jones}, {Koribalski}
  \& {Serra}}{{Kleiner} et~al.}{2017}]{kleiner}
{Kleiner} D.,  {Pimbblet} K.~A.,  {Jones} D.~H.,  {Koribalski} B.~S.,   {Serra}
  P.,  2017, \mn@doi [\mnras] {10.1093/mnras/stw3328}, \href
  {https://ui.adsabs.harvard.edu/abs/2017MNRAS.466.4692K} {466, 4692}

\bibitem[\protect\citeauthoryear{Kraljic et~al.,}{Kraljic
  et~al.}{2017}]{Kraljic_2018}
Kraljic K.,  et~al., 2017, \mn@doi [Monthly Notices of the Royal Astronomical
  Society] {10.1093/mnras/stx2638}, 474, 547

\bibitem[\protect\citeauthoryear{Kraljic, Davé  \& Pichon}{Kraljic
  et~al.}{2020}]{Kraljic_2020}
Kraljic K.,  Davé R.,   Pichon C.,  2020, \mn@doi [Monthly Notices of the
  Royal Astronomical Society] {10.1093/mnras/staa250}, 493, 362–381

\bibitem[\protect\citeauthoryear{Kraljic, Duckworth, Tojeiro, Alam, Bizyaev,
  Weijmans, Boardman  \& Lane}{Kraljic et~al.}{2021}]{Kraljic_2021}
Kraljic K.,  Duckworth C.,  Tojeiro R.,  Alam S.,  Bizyaev D.,  Weijmans A.-M.,
   Boardman N.~F.,   Lane R.~R.,  2021, \mn@doi [Monthly Notices of the Royal
  Astronomical Society] {10.1093/mnras/stab1109}, 504, 4626–4633

\bibitem[\protect\citeauthoryear{Krolewski, Ho, Chen, Chan, Tenneti, Bizyaev
  \& Kraljic}{Krolewski et~al.}{2019}]{Krolewski_2019}
Krolewski A.,  Ho S.,  Chen Y.-C.,  Chan P.~F.,  Tenneti A.,  Bizyaev D.,
  Kraljic K.,  2019, \mn@doi [The Astrophysical Journal]
  {10.3847/1538-4357/ab1010}, 876, 52

\bibitem[\protect\citeauthoryear{Kuutma, Tamm  \& Tempel}{Kuutma
  et~al.}{2017}]{Kuutma_2017}
Kuutma T.,  Tamm A.,   Tempel E.,  2017, \mn@doi [Astronomy \& Astrophysics]
  {10.1051/0004-6361/201730526}, 600, L6

\bibitem[\protect\citeauthoryear{Laigle et~al.,}{Laigle
  et~al.}{2017}]{Laigle_2017}
Laigle C.,  et~al., 2017, \mn@doi [Monthly Notices of the Royal Astronomical
  Society] {10.1093/mnras/stx3055}, 474, 5437–5458

\bibitem[\protect\citeauthoryear{Lee \& Erdogdu}{Lee \&
  Erdogdu}{2007}]{Lee_2007}
Lee J.,  Erdogdu P.,  2007, \mn@doi [The Astrophysical Journal]
  {10.1086/523351}, 671, 1248–1255

\bibitem[\protect\citeauthoryear{Libeskind et~al.,}{Libeskind
  et~al.}{2017}]{Libeskind_2017}
Libeskind N.~I.,  et~al., 2017, \mn@doi [Monthly Notices of the Royal
  Astronomical Society] {10.1093/mnras/stx1976}, 473, 1195–1217

\bibitem[\protect\citeauthoryear{Luber, Gorkom, Hess, Pisano, Fernández  \&
  Momjian}{Luber et~al.}{2019}]{luber}
Luber N.,  Gorkom J.,  Hess K.,  Pisano D.,  Fernández X.,   Momjian E.,
  2019, \mn@doi [The Astronomical Journal] {10.3847/1538-3881/ab1b6e}, 157, 254

\bibitem[\protect\citeauthoryear{Maddox et~al.,}{Maddox
  et~al.}{2021}]{Maddox_2021}
Maddox N.,  et~al., 2021, \mn@doi [Astronomy & Astrophysics]
  {10.1051/0004-6361/202039655}, 646, A35

\bibitem[\protect\citeauthoryear{Mann \& Whitney}{Mann \& Whitney}{1947}]{mwu}
Mann H.~B.,  Whitney D.~R.,  1947, \mn@doi [The Annals of Mathematical
  Statistics] {10.1214/aoms/1177730491}, 18, 50

\bibitem[\protect\citeauthoryear{{McCracken} et~al.,}{{McCracken}
  et~al.}{2012}]{McCracken2012}
{McCracken} H.~J.,  et~al., 2012, \mn@doi [\aap] {10.1051/0004-6361/201219507},
  \href {https://ui.adsabs.harvard.edu/abs/2012A&A...544A.156M} {544, A156}

\bibitem[\protect\citeauthoryear{{McMullin}, {Waters}, {Schiebel}, {Young}  \&
  {Golap}}{{McMullin} et~al.}{2007}]{casa}
{McMullin} J.~P.,  {Waters} B.,  {Schiebel} D.,  {Young} W.,   {Golap} K.,
  2007, in {Shaw} R.~A.,  {Hill} F.,   {Bell} D.~J.,  eds,  Astronomical
  Society of the Pacific Conference Series Vol. 376, Astronomical Data Analysis
  Software and Systems XVI. p.~127

\bibitem[\protect\citeauthoryear{Milnor}{Milnor}{1963}]{milnor1963morse}
Milnor J.,  1963, Stud, 51

\bibitem[\protect\citeauthoryear{Nelson et~al.,}{Nelson
  et~al.}{2019}]{Nelson_2019}
Nelson D.,  et~al., 2019, \mn@doi [Monthly Notices of the Royal Astronomical
  Society] {10.1093/mnras/stz2306}, 490, 3234–3261

\bibitem[\protect\citeauthoryear{Noether}{Noether}{1978}]{ks-two}
Noether G.,  1978, Studies in statistics, pp 39--65

\bibitem[\protect\citeauthoryear{{Pahwa} et~al.,}{{Pahwa} et~al.}{2016}]{pahwa}
{Pahwa} I.,  et~al., 2016, \mn@doi [\mnras] {10.1093/mnras/stv2930}, \href
  {https://ui.adsabs.harvard.edu/abs/2016MNRAS.457..695P} {457, 695}

\bibitem[\protect\citeauthoryear{Peebles}{Peebles}{1969}]{peebles}
Peebles P.,  1969, \mn@doi [apj] {10.1086/149876}, 155, 393

\bibitem[\protect\citeauthoryear{{Ponomareva} et~al.,}{{Ponomareva}
  et~al.}{2021}]{Ponomareva2021}
{Ponomareva} A.~A.,  et~al., 2021, \mn@doi [\mnras] {10.1093/mnras/stab2654},
  \href {https://ui.adsabs.harvard.edu/abs/2021MNRAS.508.1195P} {508, 1195}

\bibitem[\protect\citeauthoryear{{Schaap} \& {van de Weygaert}}{{Schaap} \&
  {van de Weygaert}}{2000}]{dtfe}
{Schaap} W.~E.,  {van de Weygaert} R.,  2000, Astronomy and Astrophysics, \href
  {https://ui.adsabs.harvard.edu/abs/2000A&A...363L..29S} {363, L29}

\bibitem[\protect\citeauthoryear{{Shirley} et~al.,}{{Shirley}
  et~al.}{2021}]{shirley2021help}
{Shirley} R.,  et~al., 2021, \mn@doi [\mnras] {10.1093/mnras/stab1526}, \href
  {https://ui.adsabs.harvard.edu/abs/2021MNRAS.tmp.1693S} {}

\bibitem[\protect\citeauthoryear{{Sousbie}}{{Sousbie}}{2011}]{sousbie}
{Sousbie} T.,  2011, \mn@doi [Monthly Notices of the Royal Astronomical
  Society] {10.1111/j.1365-2966.2011.18394.x}, \href
  {https://ui.adsabs.harvard.edu/abs/2011MNRAS.414..350S} {414, 350}

\bibitem[\protect\citeauthoryear{{Springel} et~al.,}{{Springel}
  et~al.}{2005}]{springel_2005}
{Springel} V.,  et~al., 2005, \mn@doi [Nature Astrophysics]
  {10.1038/nature03597}, \href
  {https://ui.adsabs.harvard.edu/abs/2005Natur.435..629S} {435, 629}

\bibitem[\protect\citeauthoryear{{Tempel}, {Stoica}  \& {Saar}}{{Tempel}
  et~al.}{2013}]{tempel2013}
{Tempel} E.,  {Stoica} R.~S.,   {Saar} E.,  2013, \mn@doi [\mnras]
  {10.1093/mnras/sts162}, \href
  {https://ui.adsabs.harvard.edu/abs/2013MNRAS.428.1827T} {428, 1827}

\bibitem[\protect\citeauthoryear{Trujillo, Carretero  \& Patiri}{Trujillo
  et~al.}{2006}]{Trujillo_2006}
Trujillo I.,  Carretero C.,   Patiri S.~G.,  2006, \mn@doi [The Astrophysical
  Journal] {10.1086/503548}, 640, L111

\bibitem[\protect\citeauthoryear{{Vulcani} et~al.,}{{Vulcani}
  et~al.}{2011}]{vulcani}
{Vulcani} B.,  et~al., 2011, \mn@doi [\mnras]
  {10.1111/j.1365-2966.2010.17904.x}, \href
  {https://ui.adsabs.harvard.edu/abs/2011MNRAS.412..246V} {412, 246}

\bibitem[\protect\citeauthoryear{{Welker}, {Devriendt}, {Dubois}, {Pichon}  \&
  {Peirani}}{{Welker} et~al.}{2014}]{welker_2014}
{Welker} C.,  {Devriendt} J.,  {Dubois} Y.,  {Pichon} C.,   {Peirani} S.,
  2014, \mn@doi [\mnras] {10.1093/mnrasl/slu106}, \href
  {https://ui.adsabs.harvard.edu/abs/2014MNRAS.445L..46W} {445, L46}

\bibitem[\protect\citeauthoryear{{Welker} et~al.,}{{Welker}
  et~al.}{2020}]{welker}
{Welker} C.,  et~al., 2020, \mn@doi [\mnras] {10.1093/mnras/stz2860}, \href
  {https://ui.adsabs.harvard.edu/abs/2020MNRAS.491.2864W} {491, 2864}

\bibitem[\protect\citeauthoryear{{White}}{{White}}{1984}]{white}
{White} S.~D.~M.,  1984, \mn@doi [\apj] {10.1086/162573}, \href
  {https://ui.adsabs.harvard.edu/abs/1984ApJ...286...38W} {286, 38}

\bibitem[\protect\citeauthoryear{Winkel, Pasquali, Kraljic, Smith, Gallazzi  \&
  Jackson}{Winkel et~al.}{2021}]{Winkel_2021}
Winkel N.,  Pasquali A.,  Kraljic K.,  Smith R.,  Gallazzi A.,   Jackson T.~M.,
   2021, \mn@doi [Monthly Notices of the Royal Astronomical Society]
  {10.1093/mnras/stab1562}, 505, 4920

\bibitem[\protect\citeauthoryear{{York} et~al.,}{{York} et~al.}{2000}]{sdss}
{York} D.~G.,  et~al., 2000, \mn@doi [\aj] {10.1086/301513}, \href
  {https://ui.adsabs.harvard.edu/abs/2000AJ....120.1579Y} {120, 1579}

\bibitem[\protect\citeauthoryear{{Yun}, {Ho}  \& {Lo}}{{Yun}
  et~al.}{1994}]{1994Natur.372..530Y}
{Yun} M.~S.,  {Ho} P.~T.~P.,   {Lo} K.~Y.,  1994, \mn@doi [\nat]
  {10.1038/372530a0}, \href
  {https://ui.adsabs.harvard.edu/abs/1994Natur.372..530Y} {372, 530}

\bibitem[\protect\citeauthoryear{{Zel'dovich}}{{Zel'dovich}}{1970}]{zeldovich-1970}
{Zel'dovich} Y.~B.,  1970, Astronomy and Astrophysics, \href
  {https://ui.adsabs.harvard.edu/abs/1970A&A.....5...84Z} {500, 13}

\bibitem[\protect\citeauthoryear{{de Lapparent}, {Geller}  \& {Huchra}}{{de
  Lapparent} et~al.}{1986}]{deLapparent-1986}
{de Lapparent} V.,  {Geller} M.~J.,   {Huchra} J.~P.,  1986, \mn@doi
  [Astrophysical Journal, Letters] {10.1086/184625}, \href
  {https://ui.adsabs.harvard.edu/abs/1986ApJ...302L...1D} {302, L1}

\makeatother
\end{thebibliography}

\bsp	
\label{lastpage}
\end{document}